\documentclass[a4paper,fleqn,usenatbib]{mnras}

\usepackage{newtxtext,newtxmath}
\usepackage{graphicx}	
\usepackage{amsmath}	
\usepackage{amssymb}	
\usepackage{newtxtext,newtxmath}
\usepackage[T1]{fontenc}
\usepackage{ae,aecompl}
\usepackage{subfigure}
\usepackage{subfloat}
\usepackage{multirow}
\usepackage{multicol}
\usepackage{lipsum}
\usepackage{dcolumn}
\usepackage{bm}

\title[Inverse Transfer of $E_M$ in a decaying MHD system]{On the Inverse Transfer of (Non) Helical Magnetic Energy in a Decaying Magnetohydrodynamic Turbulence}

\author[Kiwan Park et al.]{
Kiwan Park$^{1}$\thanks{E-mail: oz150@uni-heidelberg.de, pkiwan@gmail.com}\\
$^{1}$Center for Astronomy, Institute for Theoretical Astrophysics at University of Heidelberg 69120 Heidelberg, Germany\\}

\date{Accepted XXX. Received YYY; in original form ZZZ}

\pubyear{2017}

\begin{document}
\label{firstpage}
\pagerange{\pageref{firstpage}--\pageref{lastpage}}
\maketitle

\begin{abstract}
In our conventional understanding, large-scale magnetic fields are thought to originate from an inverse cascade in the presence of magnetic helicity, differential rotation, or a magneto-rotational instability. However, as recent simulations have given strong indications that an inverse cascade (transfer) may occur even in the absence of magnetic helicity, the physical origin of this inverse cascade is still not fully understood. We here present two simulations of freely decaying helical \& non-helical magnetohydrodynamic (MHD) turbulence. We verified the inverse transfer of helical and non-helical magnetic fields in both cases, but we found the underlying physical principles to be fundamentally different. In the former case, the helical magnetic component leads to an inverse cascade of magnetic energy. We derived a semi analytic formula for the evolution of large scale magnetic field using $\alpha$ coefficient and compared it with the simulation data. But in the latter case, the $\alpha$ effect, including other conventional dynamo theories, are not suitable to describe the inverse transfer of non-helical magnetic. To obtain a better understanding of the physics at work here, we introduced a `field structure model' based on the magnetic induction equation in the presence of inhomogeneities. This model illustrates how the curl of the electromotiveforce (EMF) leads to the build up of a large-scale magnetic field without the requirement of magnetic helicity. And we applied a Quasi Normal approximation to the inverse transfer of magnetic energy.
\end{abstract}




\section{Introduction}

Magnetic fields `$\mathbf{B}$' are ubiquitous components in the Universe, which is filled with conducting fluids (plasmas) on many different scales. The microscopic and macroscopic interactions between the magnetic field and the plasma motion lead to the amplification (dynamo) and reconnection (annihilation) of the magnetic fields. Also the evolution of magnetic fields introduces a backreaction to the plasma so that the motions of the ionized particles are constrained in reverse. But these processes, including the origin of magnetic fields, are not yet fully understood.\\

\noindent For the origin of cosmic magnetic fields,  two main hypotheses, the primordial (top-down, large to small scale) and astrophysical (bottom-up, small to large scale) model are  accepted \citep{Schmitz 2009}. The primordial model suggests magnetic fields could be generated if the conformal invariance of the electromagnetic fields was broken during the inflationary period in the early Universe \citep{Turner and Widrow 1988}. After the initial expansion, magnetic fields could be successively generated through  cosmic phase transitions such as the electroweak phase transition (EWPT) or the quantum chromodynamics transition (QCD) from quarks to hardrons \citep{Grasso and Rubinstein 2001, Subramanian 2015}. The magnitudes of the generated magnetic fields in these cosmological models are thought to have been $\mathrm{B}_0\sim 10^{-62}$G on a 1 Mpc comoving scale (during inflation), $\sim 10^{-29}$G (EWPT), and $\sim 10^{-20}$G (QCD) on a 10 Mpc scale \citep{Sigl and Olinto 1997}. The correlation lengths of these seed fields are also thought to have been limited by the scale of the particle horizon: $\sim 1-10$ cm (EWPT) and $\sim 10^4$ cm (QCD).\\

\noindent On the other hand, an astrophysical hypothesis suggests that the seed fields might have been generated by plasma effects such as the Biermann battery or the Harrison effect in the primeval astrophysical objects, posterior to the primordial inflation. The strengths of magnetic seed fields in these models are inferred to be in the range between $B_0$$\sim 10^{-21}$G \citep{Biermann 1950} and $\sim 10^{-19}$G \citep{Harrison 1970}. However, whether the seed magnetic fields have originated from primordial or astrophysical effects, the inferred strengths are too weak for the currently observed magnetic fields in our Galaxy ($\sim\mu$G). Also the inferred correlation length, which should be limited to the particle horizon at that time, is too tiny compared to that of the observed magnetic fields at present days. Thus even if the compression of magnetic fields during collapse is considered, the initial seed fields must have been subsequently amplified through a dynamo process.\\

\noindent A dynamo is essentially the redistribution of energy. It is briefly divided into two types according to the direction of the magnetic energy cascade. One is the `large scale dynamo' (LSD) where magnetic energy ($E_M$) inversely cascades toward a large scale due to the effect of (kinetic or magnetic) helicity, differential rotation ($\alpha^2\Omega$ dynamo $\rightarrow$ $\alpha^2$, $\alpha\,\Omega$ dynamo, Krause and Radler 1980, Moffatt 1980, Brandenburg 2001, Blackman and Field 2002, Park and Blackman 2012a, Park and Blackman 2012b), or the magnetorotational instability (MRI) \citep{Balbus and Hawley 1991}. The other is the `small scale dynamo' (SSD) where nonhelical magnetic energy cascades toward smaller scales so that the peak of the magnetic energy is formed between the injection and dissipation scale (\cite{Kazantsev 1968}, \cite{Kulsrud and Anderson 1992}, \cite{Brandenburg and Subramanian 2005}, references therein). The small scale dynamo has been studied during the formation of the first stars and galaxies \citep{Schleicher et al 2010}, while an $\alpha\,\Omega$ dynamo may be responsible to generate a large-scale magnetic field in the stars \citep{Charbonneau 2013} and primordial disks \citep{Latif and Schleicher 2016}.\\

\noindent Altogether the migration and increase of magnetic fields are influenced by many different factors with critical conditions that are still under debate. \citet{Kazantsev 1968}, \citet{Subramanian 1998}, \cite{Subramanian 1999} suggested that the critical magnetic Reynolds number for the dynamo should be $Re_{M, \, crit}\sim 60$. Also \citet{Haugen et al 2004} reported a critical $Re_{M, \, crit}$ in the range between $\sim$35 and $\sim$70. On the other hand, \cite{Schober et al 2012} proposed values of $Re_{M, \, crit}\sim 110$ for incompressible gas and $Re_{M, \, crit}\sim 2700$ for extremely compressible gas, i.e., for Kolmogorov and Burger turbulence, respectively. Going further, \citet{Federrath et al 2011}, \citet{Schober et al 2012}, and \citet{Schleicher et al 2013} explored the influence of Mach number and $Pr_M$ on the amplification of magnetic field in the small scale regime for the formation of primeval stars.\\

\noindent Typical theoretical and numerical dynamo models assume an external forcing source to sustain the amplification (transfer) of magnetic fields against the dissipation effects. However, it is not clear if the ever-present external driving force is necessary to form the various scales of magnetic fields observed in nature. Recently, it has been reported that the inverse transfer of energy is an intrinsic property of the MHD equations, which implies that the helicity or shear effect may not be a prerequisite for LSD \citep{Olesen 1997, Ditlevsen et al. 2004, Brandenburg et al 2015, Zrake 2014}. Olesen \& Ditlevsen et al. focused on the intrinsic properties of MHD equations and Brandenburg et al \& Zrake more focused on the simulation results. In fact \citep{Subramanian and Brandenburg 2014} showed that large scale and small scale dynamos are not entirely exclusive to each other. All of this gives us some clues to the origin of large scale magnetic fields in a quiescent astrophysical system.\\

\noindent In this paper, we present two simulations to further investigate the inverse cascade for helical and non-helical MHD turbulence along with their underlying physical properties. Instead of driving a system mechanically (kinetically driven dynamo), we drove the system magnetically and let the system decay in a free way. We first discuss the simulation results, and briefly introduce the current theories concerning the energy migration. For the decaying helical magnetic field, we will use the $\alpha^2$ dynamo theory to describe the evolution of magnetic energy and magnetic helicity on large scales. For the decaying non-helical magnetic field, we introduce a `field structure model' based on the interpretation of the curl operator in the magnetic induction equation. This model shows that magnetic energy in the presence of inhomogeneities can be transferred towards all scales regardless of the magnetic helicity ratio in principle. Finally we briefly discuss about the self-consistent dynamo process using the field structure model and simple algebra in spherical coordinate system without (with) axisymmetry, as the latter would prevent the presence of a dynamo \citep{Cowling 1934}.

\begin{figure*}
\centering{
  {
   \subfigure[]{
     \includegraphics[width=8cm]{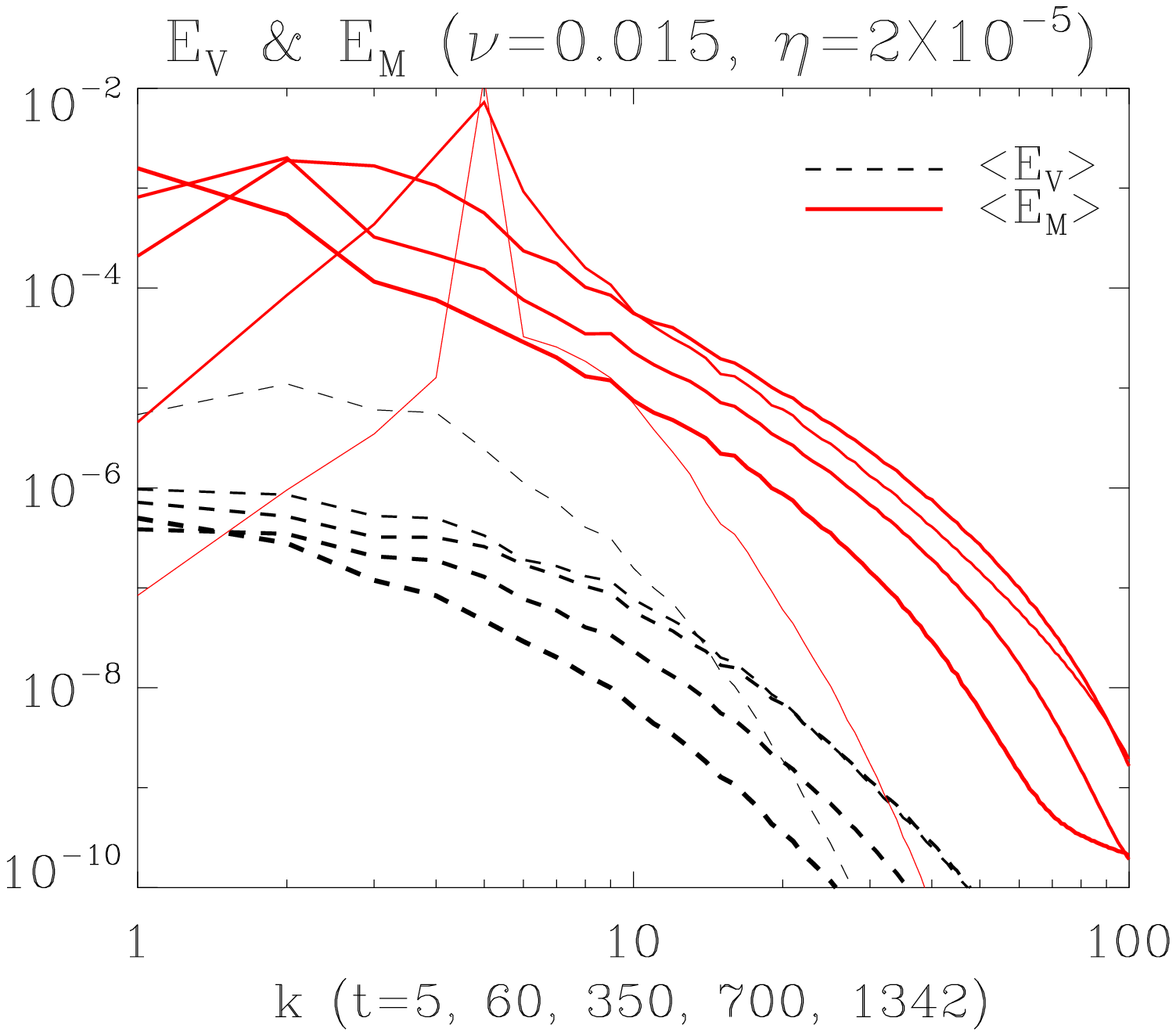}
     \label{f1}
   }\hspace{-10mm}
   \subfigure[]{
     \includegraphics[width=8cm]{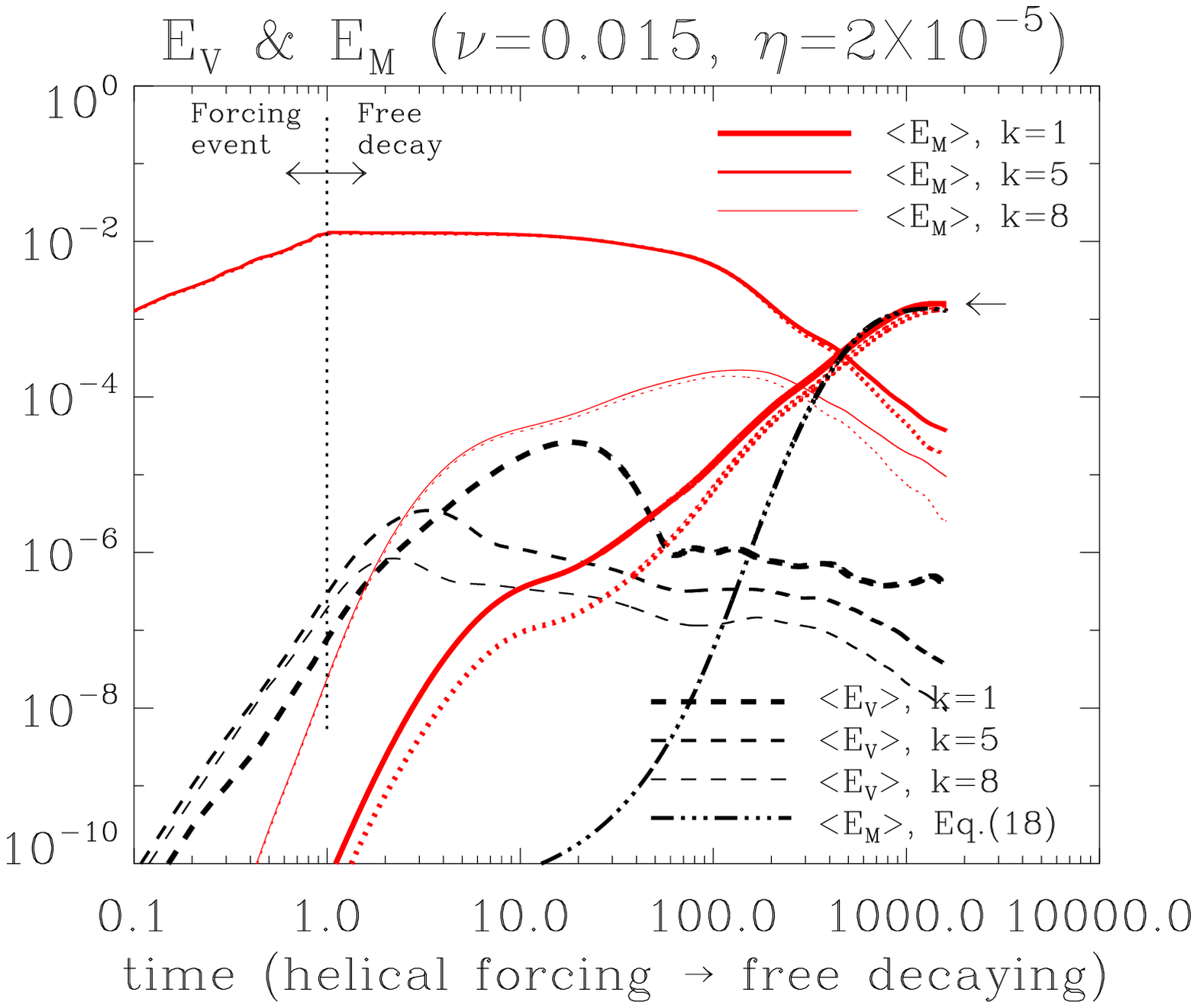}
     \label{f2}
   }\hspace{-10mm}
   \subfigure[]{
     \includegraphics[width=8cm]{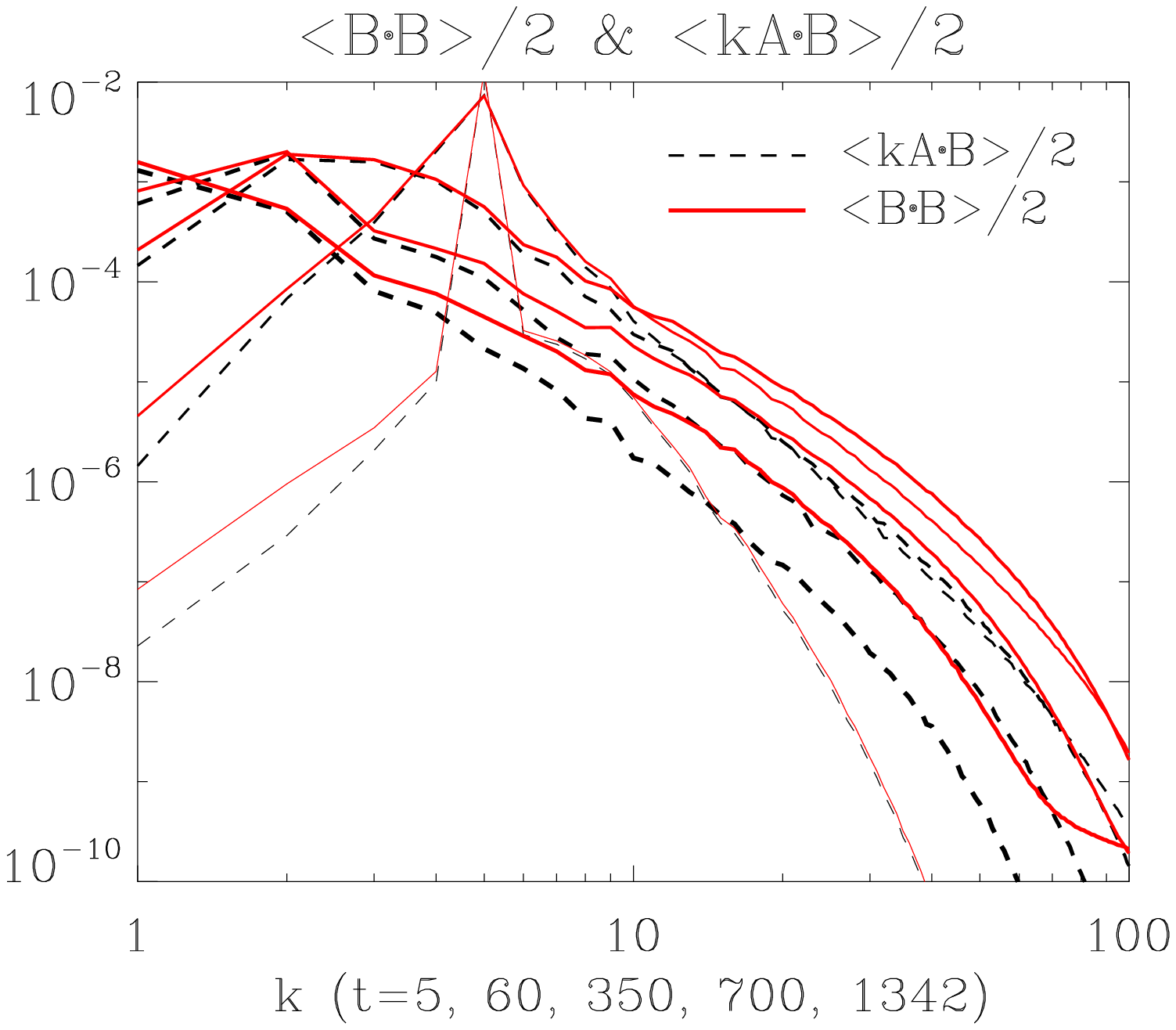}
     \label{f3}
   }\hspace{-10mm}
   \subfigure[]{
     \includegraphics[width=8cm]{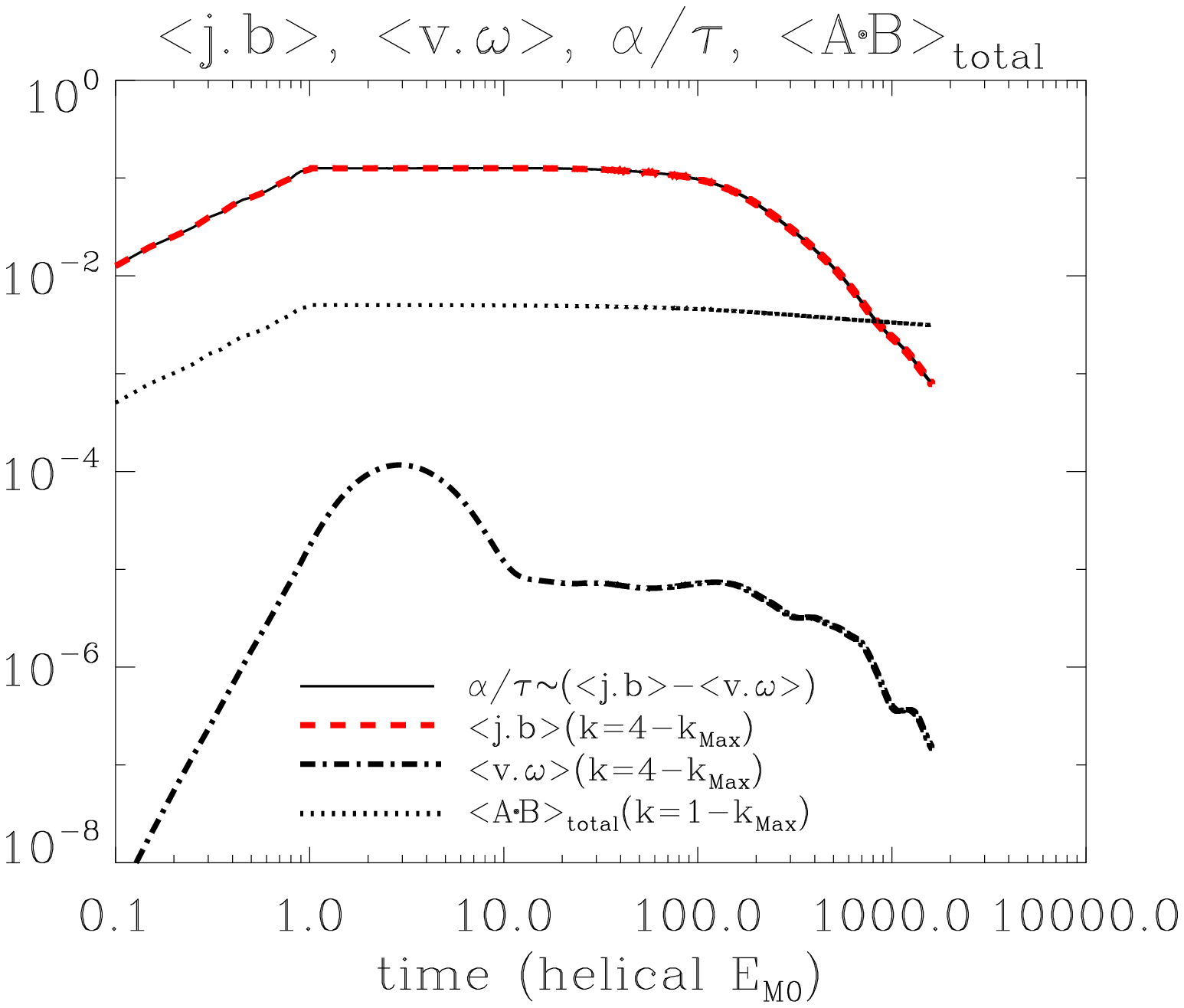}
     \label{f4}
     }
  }
\caption{(a) Spectrum of decaying $E_V$ and $E_M$ (helical + nonhelical component) after the initial helical magnetic forcing ($k=5$, $0<t\leq1$). The thinnest line indicates the earliest time (t=5), and the thickest line indicates the latest time (t=1342). (b) Temporal evolution of $E_M$ ($\langle B^2 \rangle$/2), helical $E_M$ ($\langle k\mathbf{A}\cdot \mathbf{B} \rangle$/2), and $E_V$ ($\langle u^2 \rangle$/2) at $k$=1, 5, 8. The black dot-dashed line is plotted using Eq.~(\ref{EmSolution2a}) and numerical result. (c) Spectrum of $E_M$ and its helical part. (d) Current helicity, kinetic helicity, and residual helicity ($\langle \mathbf{j}\cdot \mathbf{b}\rangle-\langle \mathbf{u}\cdot \mathbf{\omega}\rangle$) in the small scale regime ($k=4-k_{max}$). The total magnetic helicity ($k=1-k_{max}$) is also plotted. The current helicity plays a dominant role in the $\alpha$ effect, but the effect of kinetic helicity is negligibly small.} 
}
\end{figure*}

\section{Simulation and method}
For the numerical investigation we used the $\mathrm{PENCIL}$ $\mathrm{CODE}$ \citep{Brandenburg 2001}. This code solves the MHD equations for modeling  weakly compressible fluids in a periodic box $(8 \pi^3)$ with a sixth order finite difference method \citep{Vetterling 2007}. The MHD equations are coupled partial differential equations for density `$\rho$', velocity `$\mathbf{U}$', and vector potential `$\mathbf{A}$' (or magnetic field $\mathbf{B}=\nabla \times \mathbf{A}$):
\begin{eqnarray}
\frac{D \rho}{Dt}&=&-\rho {\bf \nabla} \cdot {\bf U}\label{density conservation},\label{continuity equation for pencil code}\\
\frac{D {\bf U}}{Dt}&=&-{\bf \nabla} \mathrm{ln}\, \rho + \frac{1}{\rho}(\nabla\times{\bf B})\times {\bf B}+\nu\big({\bf \nabla}^2 {\bf U}+\frac{1}{3}{\bf \nabla} {\bf \nabla} \cdot {\bf U}\big)\label{momentum equation for pencil code},\\
\frac{\partial {\bf A}}{\partial t}&=&{\bf U}{\bf \times} {\bf B} -\eta\,{\bf \nabla}{\bf \times}{\bf B}+\mathbf{f}_k,\\
\big(\Rightarrow \frac{\partial {\bf B}}{\partial t} &=& - \mathbf{U}\cdot \nabla \mathbf{B} + \mathbf{B}\cdot \nabla \mathbf{U}+\eta\,{\bf \nabla}^2{\bf B}+\nabla \times \mathbf{f}_k.\big)\nonumber
\label{magnetic induction equation for pencil code}
\end{eqnarray}
Here $D/Dt(=\partial / \partial t + {\bf U} \cdot {\bf \nabla}$)\footnote{`$\mathbf{U}$' includes the large and small scales. But in theory `$\overline{\mathbf{U}}$' in large scale (k=1) is approximately neglected after Galilean transformation. Instead a small letter `$\mathbf{u}$' is used.} is the Lagrangian time derivative to be calculated following the trajectory of the fluid motion. Magnetic diffusivity and kinematic viscosity are represented by $\eta$ (=$c^2/4\pi \sigma$, $c$: speed of light, $\sigma$: conductivity) and $\nu$ (=$\mu/\rho$, $\mu$: viscosity) respectively. The velocity is in units of the sound speed $c_s$, and the magnetic field is given in units of $(\rho_0\,\mu_0)^{1/2}c_s$ ($[B]=\sqrt{\rho_0\,[\mu_0]}[v]$ from the magnetic energy density $E_M$ ($\equiv B^2/2\mu_0$) and the kinetic energy density $E_{V}$ ($\equiv \rho_0 U^2/2$)). In addition, $\mu_0$ represents the magnetic permeability, and $\rho_0$  the initial density. Note that $\rho$ is approximately $\rho_0$ in the weakly compressible simulations. The forcing function ${\bf f}(x,t)$ is represented by $N\,{\bf f}_k(t)\, exp\,[i\,{\bf k}_f(t)\cdot {\bf x}+i\phi(t)]$ ($N$: normalization factor, ${\bf f}_k(t)$: forcing amplitude, ${\bf k}(t)$: forcing wave number), where ${\bf f}_k(t)$ is
\begin{eqnarray}
{\bf f}_k(t)=\frac{i\mathbf{k}(t)\times (\mathbf{k}(t)\times \mathbf{e})-\xi |k(t)|(\mathbf{k}(t)\times \mathbf{e})}{k(t)^2\sqrt{1+\xi^2}\sqrt{1-(\mathbf{k}(t)\cdot \mathbf{e})^2/k(t)^2}}.\nonumber
\label{forcing ampliitude fk}
\end{eqnarray}
Here `$\mathbf{e}$' is an arbitrary unit vector, and `$\xi$' denotes the helicity ratio. For example if `$\xi$' is `1', we get the relation of a fully helical field: $i\mathbf{k}\times \mathbf{f}_k=|k|\mathbf{f}_k$. On the contrary if `$\xi$' is `0', $i\mathbf{k}\times \mathbf{f}_k$ is not proportional to $\mathbf{f}_k$, i.e., $i\mathbf{k}\times \mathbf{f}_k\nsim\mathbf{f}_k$. With $\xi$ in the range of (0, 1) the forcing function generates a partially helical force. The constants $c_s$, $\mu_0$, and $\rho_0$ are set to be `1' . This choice defines the unit system employed in the simulations.\\

\noindent The injection scale (forcing wavenumber) in the simulation is $k_f=5$. The kinematic viscosity $\nu$ and the magnetic diffusivity $\eta$ are $0.015$ and $2\times 10^{-5}$ respectively. And the system was forced magnetically \citep{Park and Blackman 2012b} with $f_k=0.02$ with the resolution of $288^3$. We have a magnetic Prandtl number of $Pr_M$=$\nu/\eta=750$, which we take as a good approximation for the large values $Pr_M\gg 1$ expected in the early universe (Kulsrud 1999, Schekochihin et al 2002, Schober et al. 2012, and references there in). Turning on and off the forcing function ($0<t<1$, simulation time unit), we imitate an ephemeral event which had driven a celestial (MHD) system in the past.\\

\begin{figure*}
\centering{
  {
   \subfigure[]{
     \includegraphics[width=8cm]{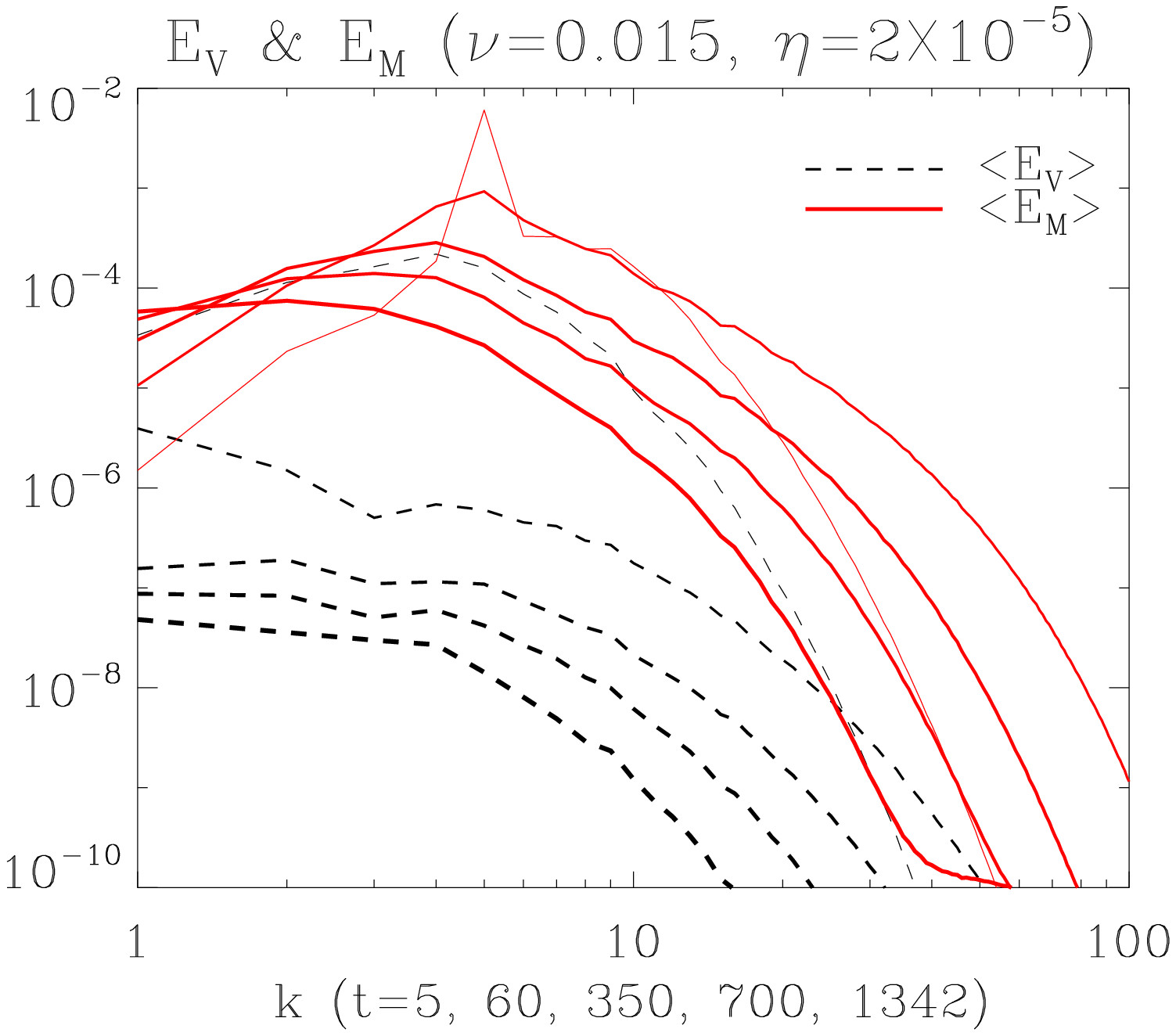}
     \label{f5}
   }\hspace{-10mm}
   \subfigure[]{
     \includegraphics[width=8cm]{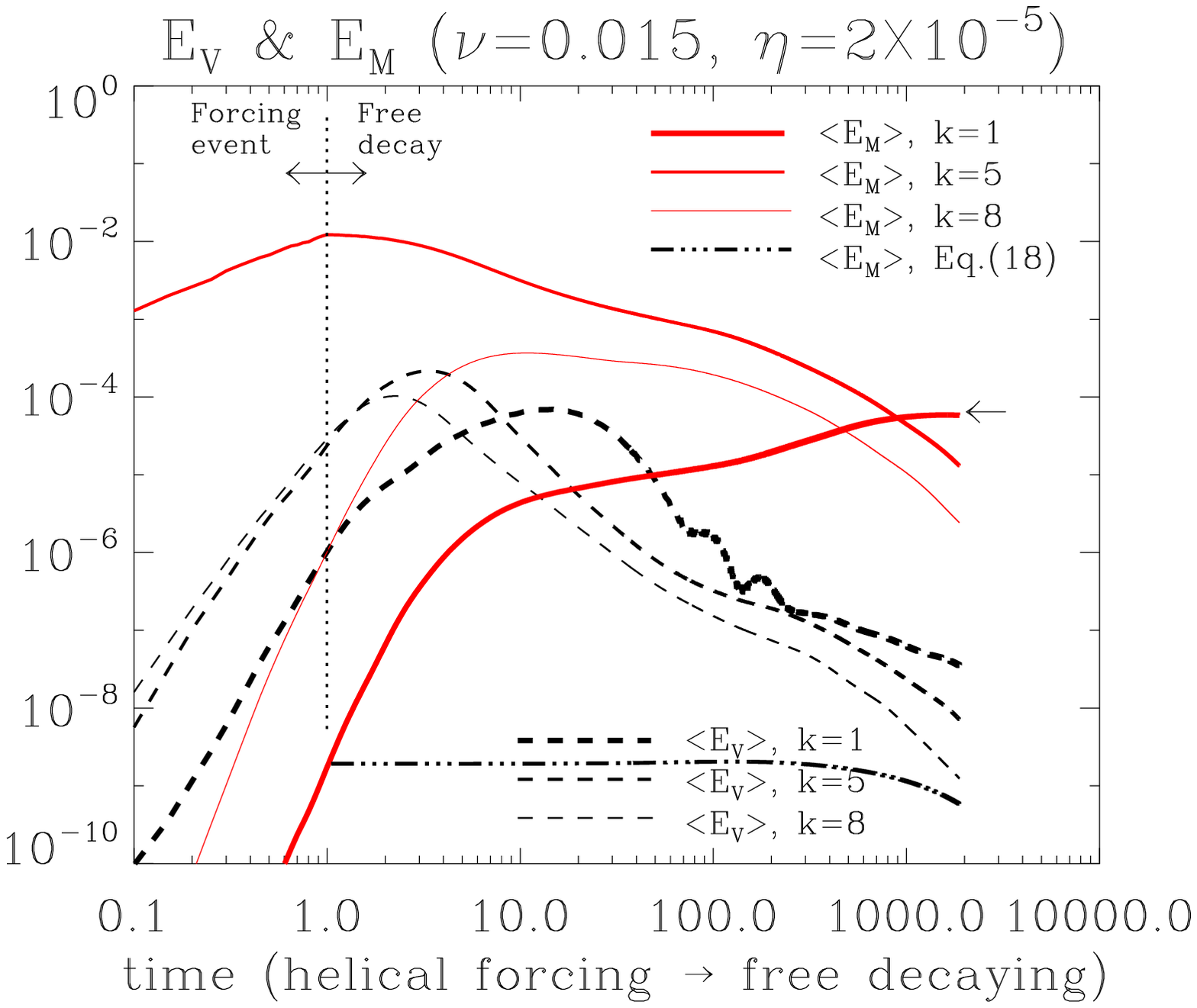}
     \label{f6}
     }\hspace{-10mm}
   \subfigure[]{
     \includegraphics[width=8cm]{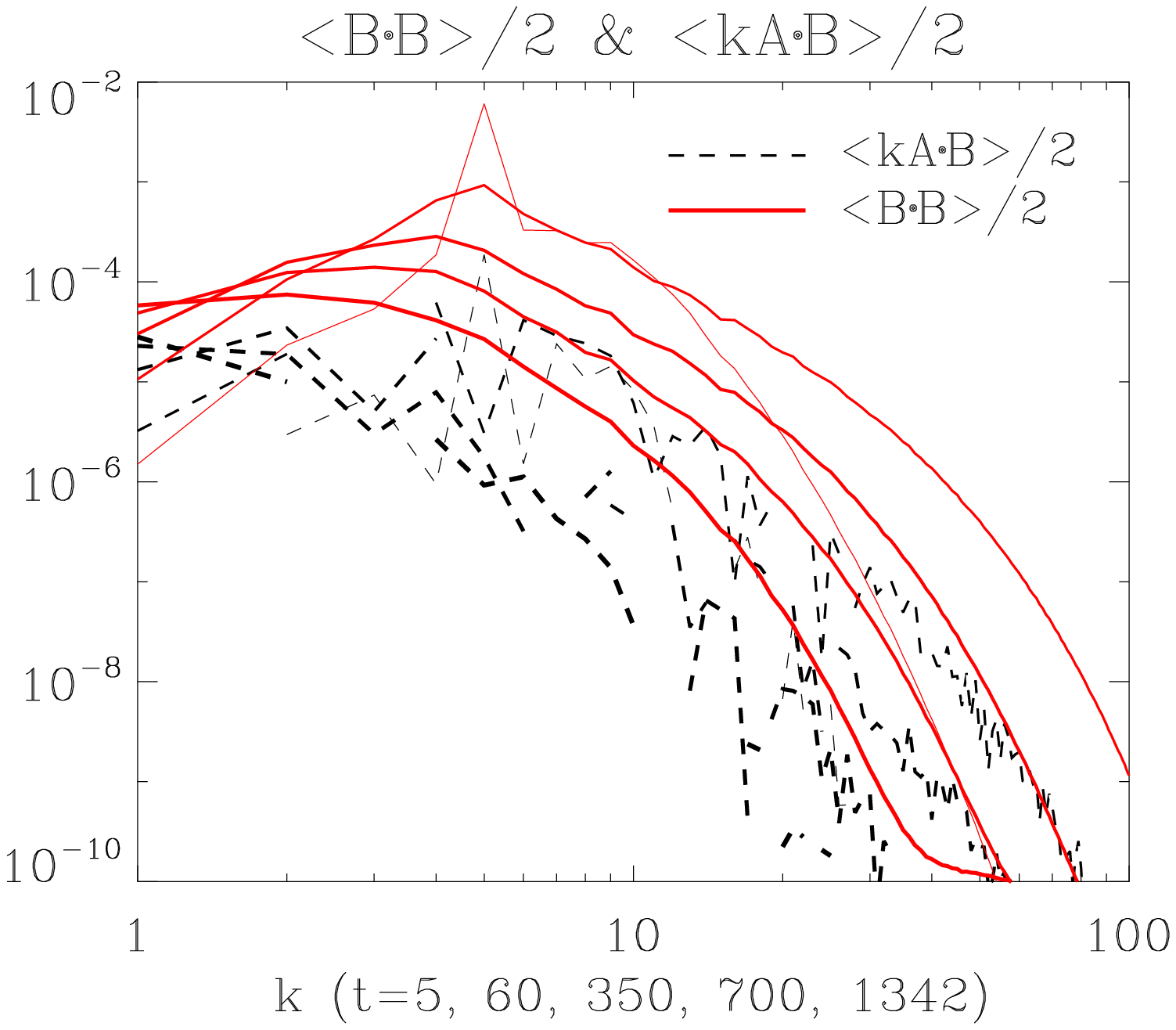}
     \label{f7}
     }\hspace{-10mm}
   \subfigure[]{
     \includegraphics[width=8cm]{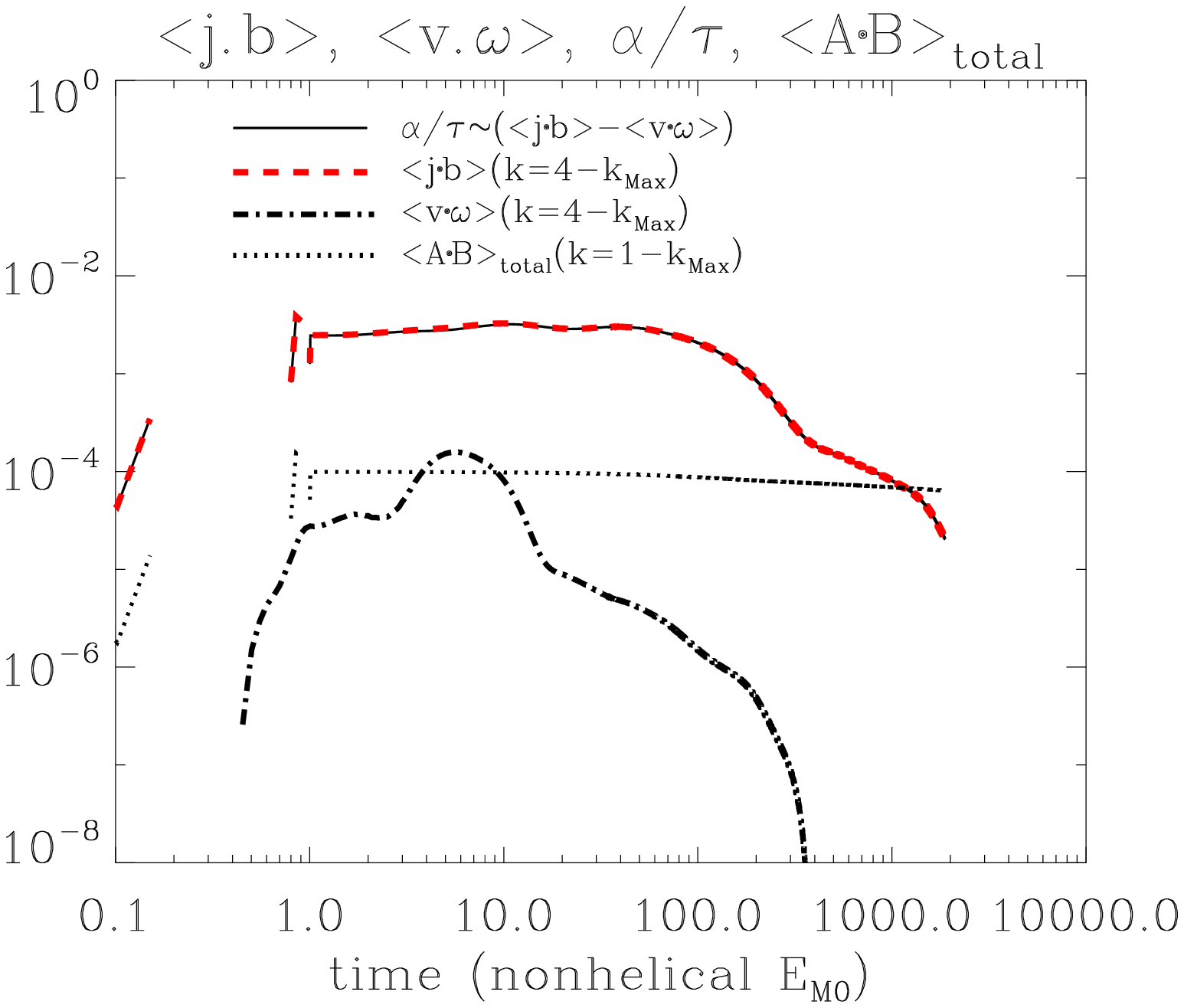}
     \label{f8}
     }
  }
\caption{Decay after nonhelical forcing. All conditions in these plots are the same as those of Fig.~\ref{f1}-\ref{f4} except the magnetic helicity ratio (nonhelical forcing). The dotted-dashed line in Fig.~\ref{f6} was made with Eq.~(\ref{EmSolution2a}) and simulation data.}
}
\end{figure*}

\begin{figure*}
\centering{
  {
   \subfigure[]{
     \includegraphics[width=8cm]{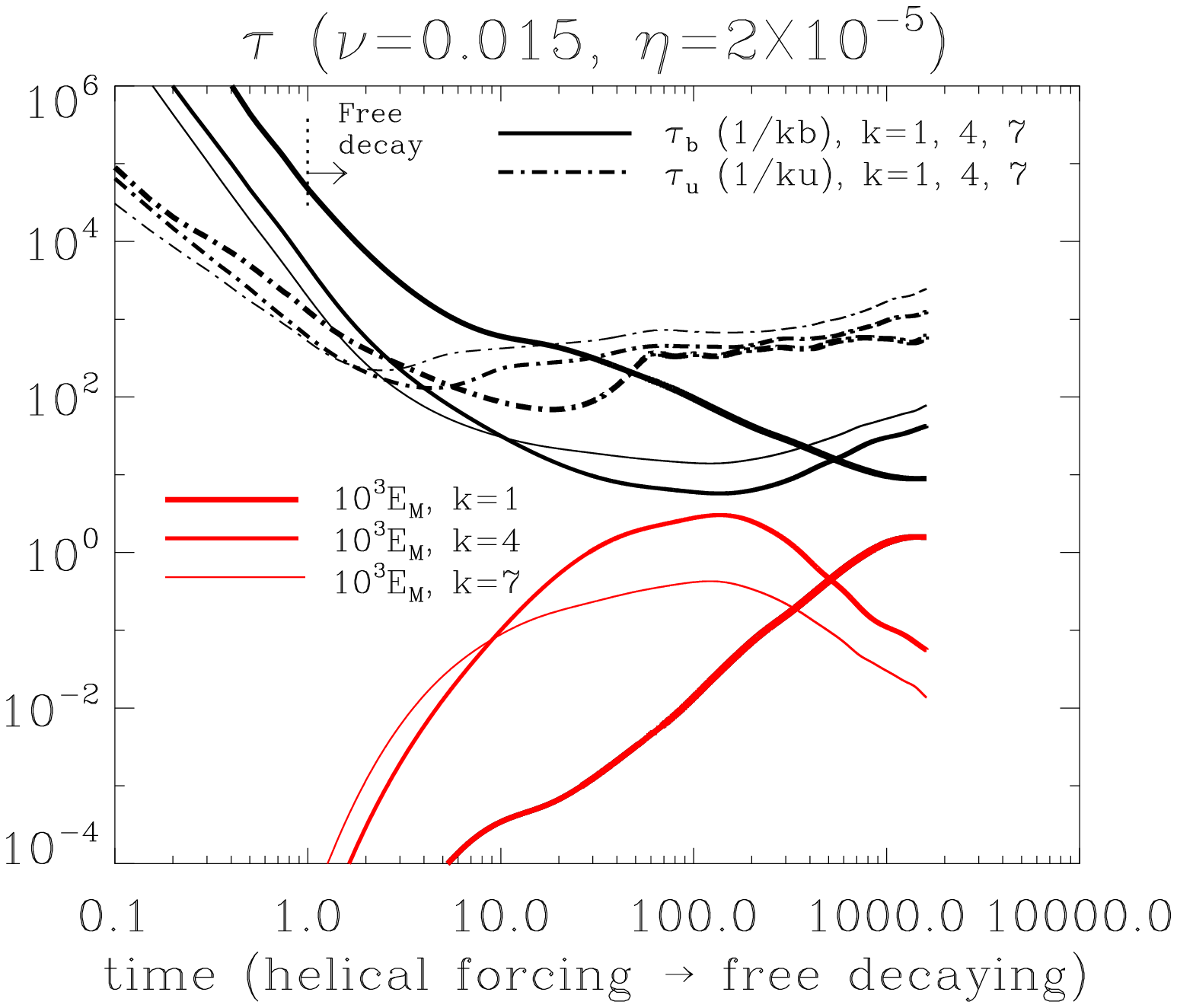}
     \label{Tauhelical}
   }\hspace{-10mm}
   \subfigure[]{
     \includegraphics[width=8cm]{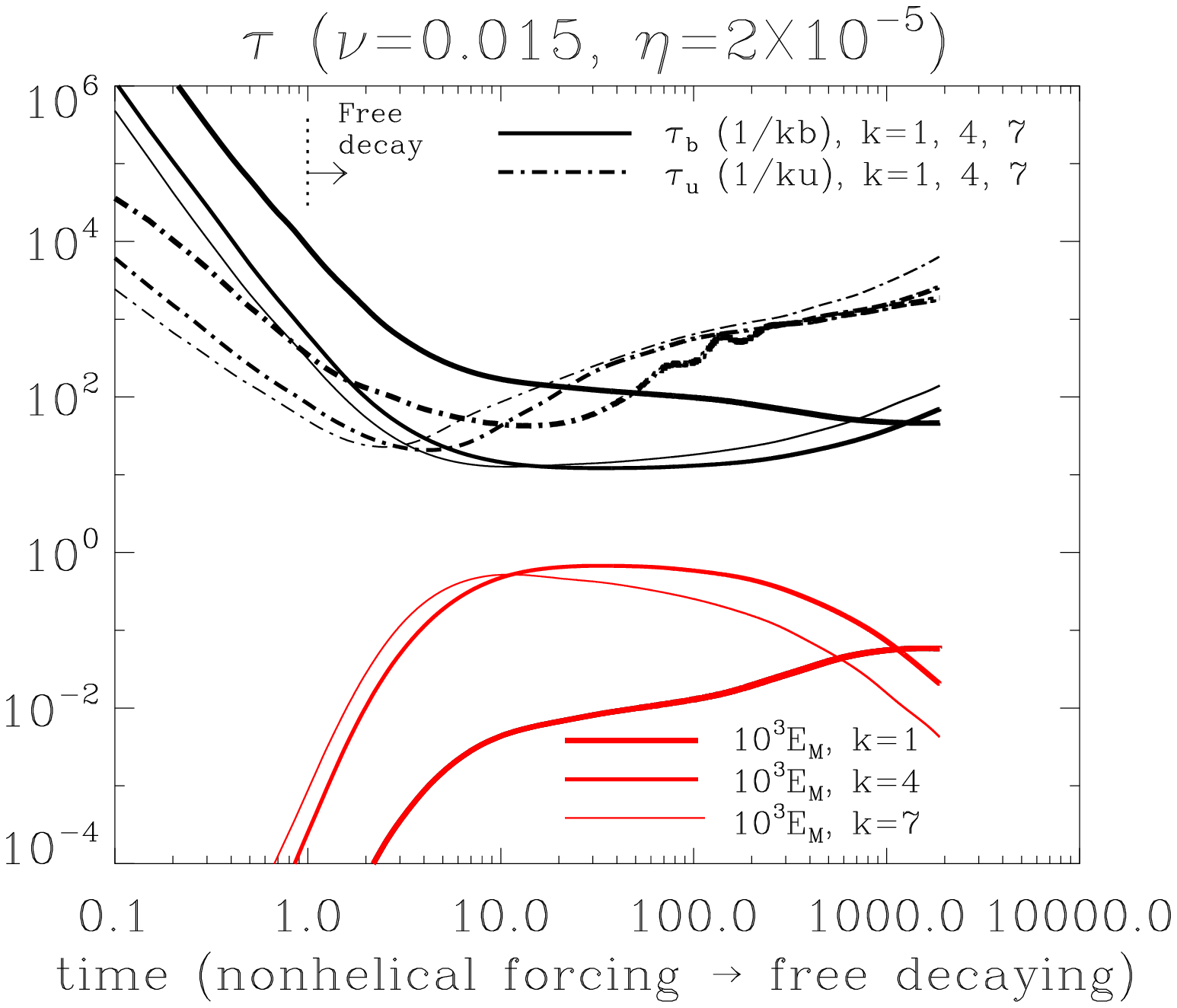}
     \label{TauNonhelical}
     }\hspace{-10mm}
  }
\caption{(a) The lines in the upper part indicate the magnetic (solid line) and kinetic (dash-dot line) eddy turnover time for k=1 (thickest), 4, 7 (thinnest). The red solid lines in the lower part show the evolving $E_M (\times10^3)$. The injection scale $k=5$ in this system was driven by fully helical magnetic energy for $0<t<1$.
(b) The plot includes the kinetic and magnetic eddy turnover time for the decaying nonhelical system.}
}
\end{figure*}

\section{Numerical Results}
Fig.~\ref{f1} shows the freely decaying spectrum of the kinetic energy density $E_V$ (black dashed line) and the magnetic energy density $E_M$ (red solid line) after the initial helical forcing (magnetic forcing) at $k=5$ ($0\leq t\leq1$). The increasing thickness of the lines corresponds to the lapse of time from $t=5$ (thinnest line) to $t=1342$ (thickest line). The plot shows that the overall profiles of $E_V$ and $E_M$ decay, with the exception of the large scale component of the magnetic energy. The inversely cascading peak of the large scale $E_M$ is clearly observed. Fig.~\ref{f2} shows the temporal profile of $E_M$ ($\langle B^2\rangle /2$, helical + nonhelical component, red solid line. The symbol `$\langle \cdot\rangle$' indicates the average (mean) over large scale ($k$=1).), helical $E_M$ ($k\langle \mathbf{A}\cdot \mathbf{B}\rangle /2$, $\langle \mathbf{A}\cdot \mathbf{B}\rangle$: magnetic helicity, red dotted line), and $E_V$ ($\langle U^2\rangle /2$, black dashed line) at $k=1, 5, 8$ (representing large scale, injection scale, and small scale, respectively). The plot clearly shows the growth of $E_V$ and $E_M$ on all scales before they decay. The positive current helicity ($\equiv\langle \mathbf{j}\cdot \mathbf{b}\rangle=k^2\langle \mathbf{a}\cdot \mathbf{b}\rangle$, $\mathbf{j}=\nabla \times \mathbf{b}=k\,\mathbf{b}$) supplied by the initial forcing cascades the magnetic energy from small to large scales through the $\alpha$ effect \citep{Moffatt 1980, Krause and Radler 1980, Biskamp 2008}. The evolution of $E_M$ at $k=1$ implies that the large scale magnetic energy will also decay eventually. The onset and evolution of large scale $E_M$ and $E_V$ lag behind those of the energy densities on the small scale, which is mostly due to the scale-dependent eddy turnover time or energy cascade time. Also to check the $\alpha$ effect, the analytic result Eq.~\ref{EmSolution2a} with the simulation data for $\alpha$, $\beta$ coefficient is included.\\

\noindent The plots show some features in the field evolution. First, the faster evolution of $E_V$ is one of the typical properties of large $Pr_M$ systems. Second, the migration of magnetic energy is bi-directional whether the field is helical or not. A careful look at Figs.~\ref{f2}, \ref{f3} shows that helical (nonhelical) large and small scale $E_M$ grow before decay. This apparent bi-directional cascade of fully helical magnetic field may look contradictory to the conventional dynamo theory, which clearly explains the inverse cascade of helical field.\\

\begin{figure*}
\centering{
  {
   \subfigure[Decay with initial helicity]{
     \includegraphics[width=6.3cm]{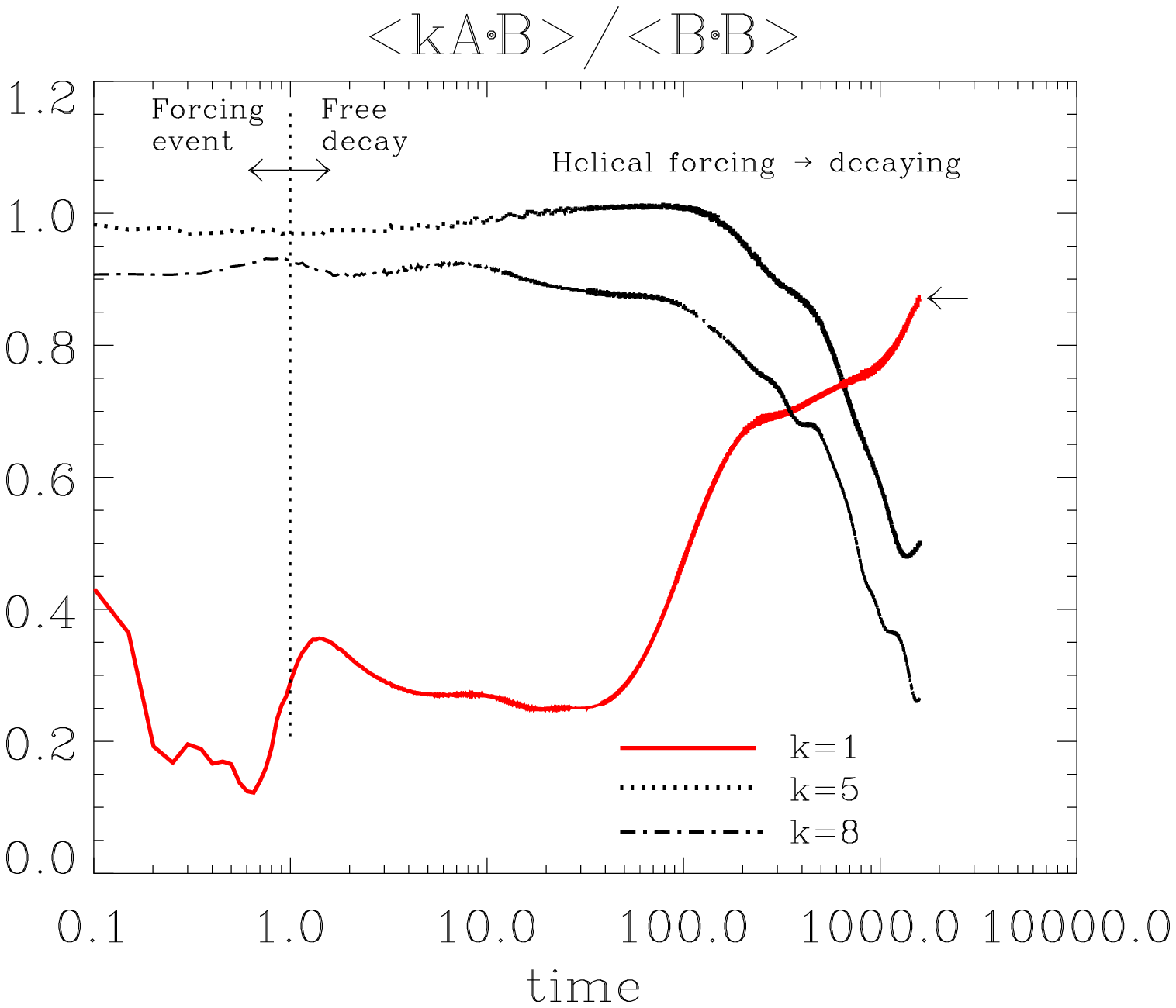}
     \label{f9}
   }\hspace{-12mm}
   \subfigure[Decay without initial helicity]{
     \includegraphics[width=6.3cm]{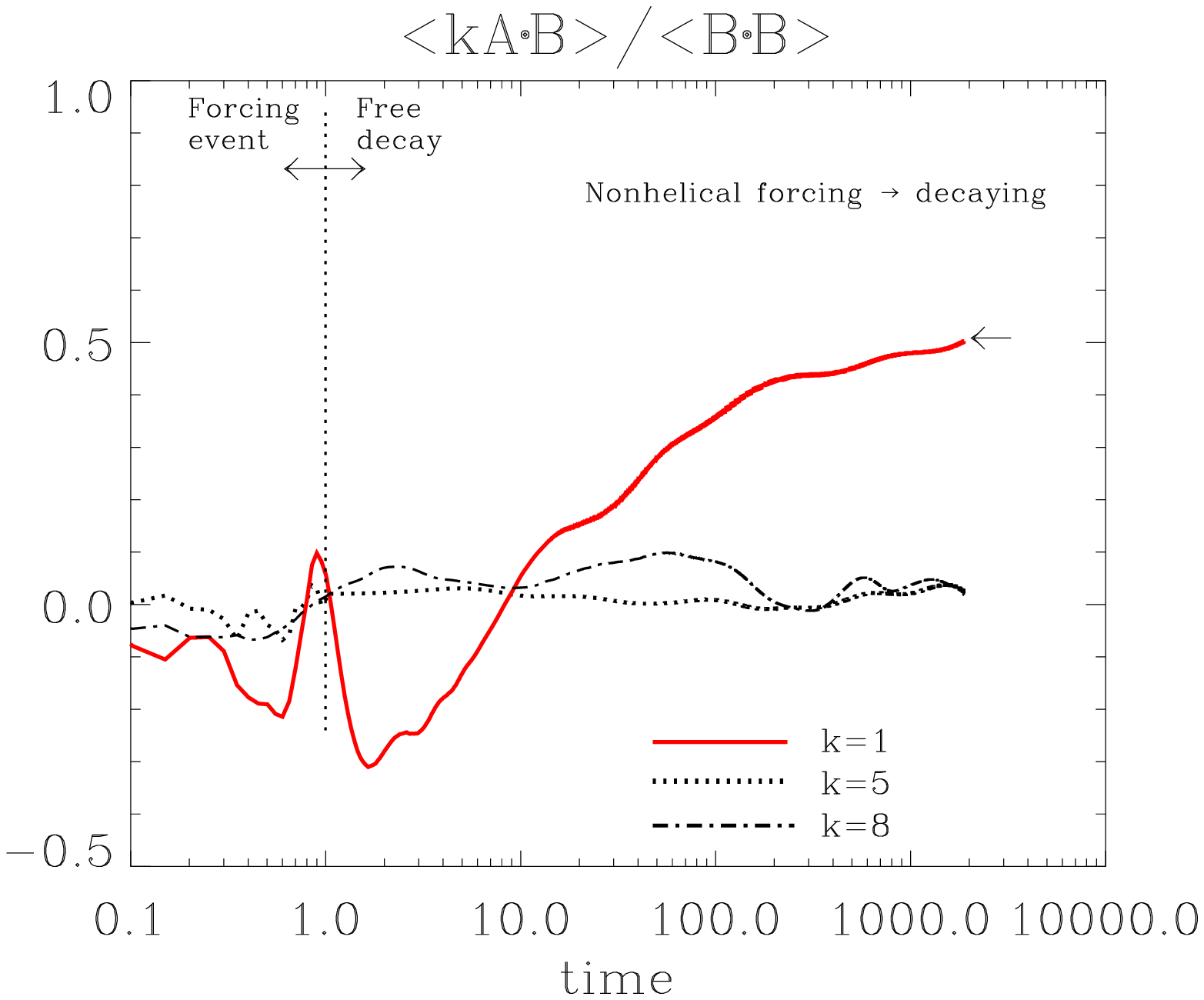}
     \label{f10}
   }\hspace{-12mm}
   \subfigure[$E_{ML,\, NH}/E_{ML,\, H}$, $\alpha_{NH}/\alpha_H$]{
     \includegraphics[width=6.3cm]{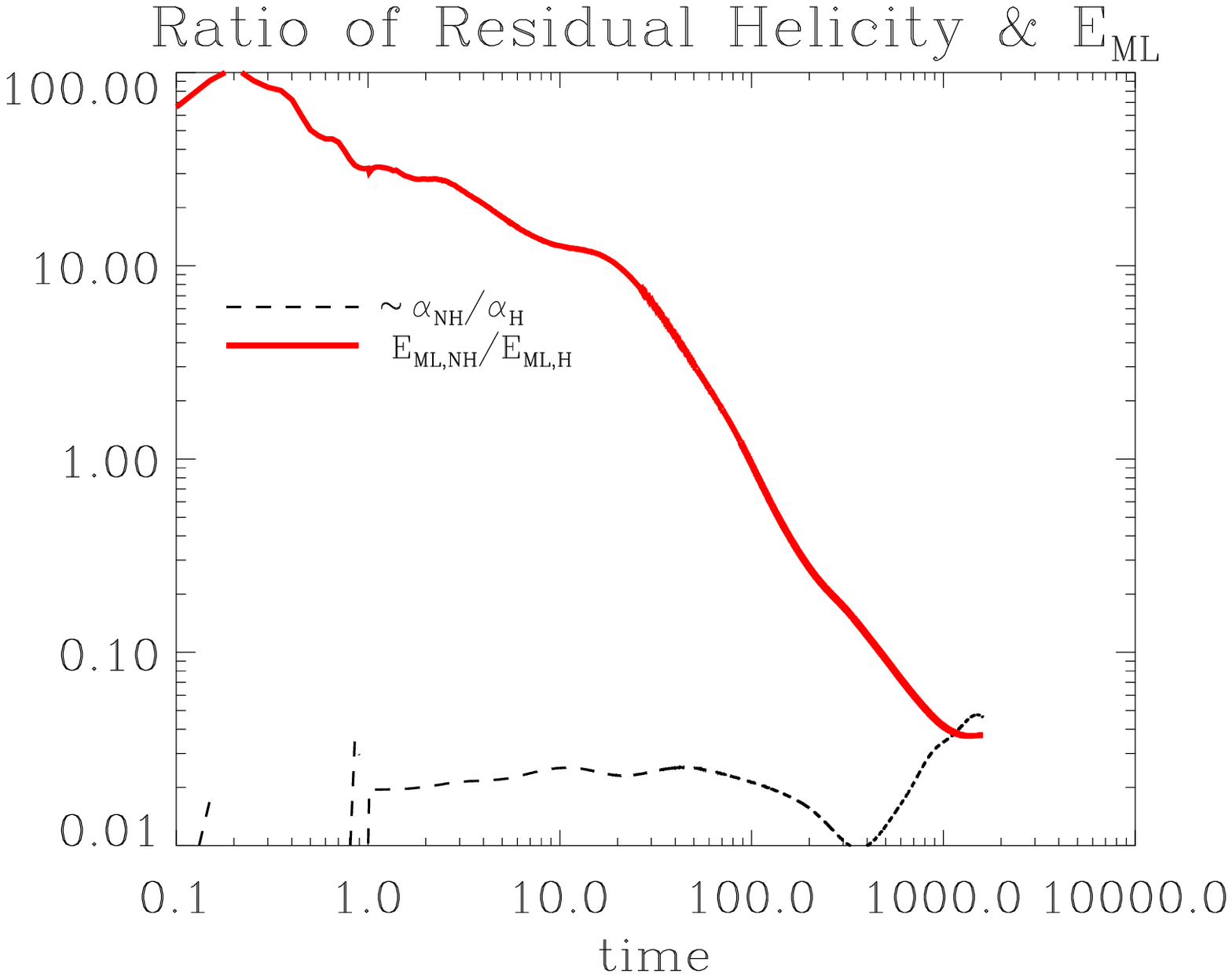}
     \label{f11}
     }
  }
\caption{(a), (b) Helicity ratio $\langle k\mathbf{A} \cdot \mathbf{B}\rangle$/$\langle \mathbf{B} \cdot \mathbf{B}\rangle$ at $k=1, 5, 8$. In (b) the growing helical field in the large scale is naturally generated. (c) For $t < \sim100$ large scale nonhelical $E_M$ is larger than helical $E_M$, i.e., $E_{ML, NH}>E_{ML, H}$.}
}
\end{figure*}

\noindent However, since the helical magnetic field can also generate nonhelical magnetic field through the induction equation $\partial \mathbf{B}/\partial t \sim \nabla \times \langle\mathbf{u}\times \mathbf{b}\rangle\sim-\mathbf{u}\cdot \nabla \mathbf{b}+\mathbf{b}\cdot \nabla \mathbf{u}$, the forward cascade (induction) of magnetic field does not conflict with the LSD. Third, the magnetic helicity ratio $f_h$ \footnote{$f_h\equiv k\,\mathbf{a}\cdot \mathbf{b}/b^2$, i.e., the ratio of helical energy to the total energy. Also we use $(b^2-k\,\mathbf{a}\cdot \mathbf{b})/2$ for the nonhelical energy. And for the large scale (k=1), the dimensions of magnetic energy, magnetic and current helicity are the same.} of the small scale $E_M$ decreases with time but that of the large scale $E_M$ increases. The increasing magnetic helicity ratio of the large scale component reduces the effect of the Lorentz force $\langle \mathbf{J}\times \mathbf{B}\rangle$ $(\rightarrow \sim k\langle \mathbf{B} \times \mathbf{B} \rangle)$ on the momentum equation. As a result, for $t>\sim60$ the large scale $E_V$ does not change much in spite of a growing large-scale magnetic field. At the same time, the small scale $E_V$ keeps rather constant in spite of decreasing $E_M$ ($\sim 100 < t < \sim 400$). Moreover, since the magnetic helicity is a better conserved variable because of its smaller order of the spatial derivative, the decreasing helical component of the small scale magnetic energy clearly indicates its inverse cascade. Fourth, the growth of helical and nonhelical magnetic energy in the small scale regime shows that the generation of a non-helical field in a fully helically driven turbulent system is a natural phenomenon. The generation of helical field from the nonhelical field is also a natural phenomenon. We will see that the magnetic helicity ratio on large scale also increases without a helical energy source (Fig.~\ref{f10}).\\

\noindent Fig.~\ref{f4} shows the time evolution of current helicity (red dashed line), kinetic helicity ($\langle \textbf{u} \cdot \omega \rangle$, $\omega=\nabla \times \mathbf{u}$: vorticity, black dot-dashed line), and residual helicity (`$\langle \textbf{j}\cdot \textbf{b}\rangle$ -$\langle \textbf{u}\cdot \omega\rangle$', black solid line, $\tau$: eddy turnover time or energy cascade time, \cite{Biskamp 2008}) on small scales ($k=4-k_{max}$). The $\alpha$ coefficient is derived as `$1/3\int^{\tau}\big(\langle {\bf j}\cdot {\bf b}\rangle-\langle {\bf u}\cdot {\bf\omega}\rangle\big ) dt$', which is composed of the small scale kinetic helicity $\langle {\bf u}\cdot {\bf\omega}\rangle$ and the current helicity $\langle {\bf j}\cdot {\bf b}\rangle$. This coefficient indicates that the small scale quantities transfer the magnetic energy toward the large scale in an energy cascade time `$\tau$'. The residual helicity, `$\langle \textbf{j}\cdot \textbf{b}\rangle$ -$\langle \textbf{u}\cdot \omega\rangle$', can be approximately represented by $\sim\alpha/\tau$\footnote{`$\tau$' here means energy correlation time, which is not yet exactly defined. In this paper we assume the `energy correlation time' is approximately same as `energy cascade time' and `eddy turnover time' for simplicity.}. If $\tau$ is related to the plasma motion, we define $\tau_u\sim 1/ku$. But if $\tau$ is related to the magnetic energy, we define $\tau_b\sim 1/kb$. Also the dissipation time scale can be defined like $\tau_{\nu}\sim1/\nu k^2$ and $\tau_{\eta}\sim1/\eta k^2$. In the plot the total magnetic helicity ($\langle \textbf{A}\cdot \textbf{B}\rangle$, $k=1-k_{max}$, black dotted line) is included for comparison. The profiles of residual and current helicity are almost overlapping during decay, which means that the influence of kinetic helicity is ignorably small. The current helicity is conserved until $t\sim 80$, and then begins to decay. In contrast the total magnetic helicity is well conserved during the whole decaying process. For $\sim10 < t < \sim100$, the kinetic helicity remains relatively constant in spite of decreasing $E_V$. The growth of the kinetic helicity is caused by various factors. As the evolution equation of the kinetic helicity shows
\begin{eqnarray}
\frac{d}{dt}\langle \mathbf{u}\cdot\omega \rangle &=& - \nabla\cdot \big[\big(p-\frac{1}{2}v^2\big)\omega\big]+\overbrace{\omega\cdot \langle \mathbf{j}\times \mathbf{b} \rangle + \mathbf{u}\cdot \nabla \times\langle \mathbf{j}\times \mathbf{b} \rangle}^{source}\nonumber\\
&&-2 \nu \langle(\partial_ju_i) (\partial_j\omega_i)\rangle + \nu \nabla^2 \langle \mathbf{u}\cdot\omega \rangle,
\label{kinetic helicity eqs}
\end{eqnarray}
we can infer that  increasing $\langle \mathbf{j}\times \mathbf{b} \rangle$, i.e., decreasing magnetic helicity, may contribute to this phenomenon. (See the small scale magnetic helicity ratio in Fig.~\ref{f9}. The overall magnetic helicity in the small scale regime decays for $t>\sim60$). The fact that a non-helical magnetic field can generate a helical velocity field is a particularly interesting issue in this respect. \\

\noindent Figs.~\ref{f5}-\ref{f8} show the evolution of the energy densities $E_V$ and $E_M$ in Fourier and real space. All conditions of this simulation are the same as in Figs.~\ref{f1}-\ref{f4}, except the helicity ratio of the initial driving force. For $0\leq t\leq1$ the system was driven by a random non-helical force at $k=5$, and then the system was let to decay. Fig.~\ref{f5} shows that $E_V$ and $E_M$ on all scales except the large scale seem to monotonically decay like in the helical forcing case. However as Fig.~\ref{f6} shows $E_V$ and $E_M$ initially keep growing (except on the injection scale) after the initial forcing stopped. Since the system is not forced, we can infer that the growth of the magnetic energy on small and large scales  is caused by the transfer of non-helical magnetic energy from the injection scale. This backward transfer of magnetic energy without helicity or shear effect contradicts the conventional helical dynamo theory or MRI theory. However, the mathematical meaning of $\nabla \times \langle \mathbf{u}\times \mathbf{b}\rangle$ and its Fourier representation give us some clues to the (inverse) transfer of general magnetic energy, which we will explore in section 3.2 in further detail. To begin with, we need to check if a helical magnetic field component might exist in the system and may have transferred the energy inversely. Without a source of the helical magnetic field, the helical component can be generated naturally in a statistically equilibrium state. This naturally generated helical magnetic component is related to the (quasi) conservation of system variables. The equilibrium state relation for the conserved energy and magnetic helicity is given as \citep{Biskamp 2008}
\begin{eqnarray}
\delta \bigg(\int \frac{1}{2}(U^2+B^2)\,dV-\frac{1}{2}\lambda\int \mathbf{A}\cdot \mathbf{B}\,dV\bigg)=0.
\label{Gen of A.B1}
\end{eqnarray}
And the variation of the vector potential $\mathbf{A}$ in this equation leads to the relation for the helical field like below:
\begin{eqnarray}
(\nabla \times \nabla \times \mathbf{A}) \cdot \delta \mathbf{A} - \lambda \nabla \times \mathbf{A}  \cdot \delta \mathbf{A} = 0
\Rightarrow \nabla \times \mathbf{B} = \lambda \mathbf{B}.
\label{Gen of A.B2}
\end{eqnarray}
Indeed the naturally generated magnetic helicity is shown in Figs.~\ref{f8}, \ref{f10}. However as the logarithmic plot of magnetic helicity in Fig.~\ref{f8} shows, the average of the generated helical field during the non-helical forcing is negligibly small. And by the end of the forcing, a small but nontrivial magnetic helicity is generated, which implies that the decaying mode in a large $Pr_M$ system is closer to the statistically equilibrium state rather than the driven mode. Nevertheless, as the comparison of Fig.~\ref{f4} and Fig.~\ref{f8} shows, its strength is much smaller than in the helically driven case ($<\sim1\%$). Therefore, the increasing helical component on the large scale shown in Fig.~\ref{f10} is not due to the inverse cascade of small scale magnetic energy. To demonstrate this, Eq.~(\ref{EmSolution2a}), calculated with the simulation data, was drawn and included in Fig.~\ref{f6}. The plot clearly shows that the $\alpha$ effect is negligibly small. As a result it can be concluded that the effect of $\alpha$ coefficient practically does not exist. The inverse transfer of nonhelical magnetic energy in a decaying MHD system deviates from the typical dynamo mechanism but reflects a fundamental principle of energy migration in a magnetized plasma system without helicity.\\

\noindent On the other hand the spectrum shown in Fig.~\ref{f1} and Fig.~\ref{f5} are quite similar. But the evolving profile of large scale $E_M$ in each case is quite different. The $\alpha$ effect in Fig.~\ref{f1} boosts the growth of $E_M$ at later time, but the growth of large scale $E_M$ in Fig.~\ref{f5} is rather suppressed. This can be verified by Fig.~\ref{f2} and Fig.~\ref{f6}.\\

\noindent Fig.~\ref{Tauhelical}, \ref{TauNonhelical} are to show the evolution of eddy turnover time. We will discuss their meanings later.\\

\noindent Fig.~\ref{f9}, \ref{f10} are to show the relation between the energy transfer and `$\tau$'. Fig.~\ref{f9} includes the evolving $\tau(k)$ and $10^3\times E_M(k)$ of $k=1,\,4,\,7$ in the decaying helical system. And Fig.~\ref{f10} includes $\tau(k)$ and $10^3E_M(k)$ of $k=1,\,4,\,7$ in the decaying nonhelical system. A careful look tells us that the change of $\tau_u (\sim 1/ku)$ precedes that of $\tau_b (\sim 1/kb)$. $E_M$ of $k=7$ with the smallest $\tau_u$ grows fastest initially, followed by $E_M$ of $k=4$ and $k=1$. The field evolution of the decaying nonhelical system shows a similar trend. $\tau_u \sim 1/ku$ is related to the local energy transfer from a magnetic eddy to another neighboring magnetic one, and the direction of energy transfer is perpendicular to the magnetic eddy, i.e., $k_\bot$.\\

\noindent Fig.~\ref{f9} includes the evolving magnetic helicity ratio for $k=1$ (red solid line), $k=5$ (black dotted line), and $k=8$ (black dot-dashed line) for the helically driven system. Fig.~\ref{f10}  includes the same contents, but they are for the non-helically driven system. In case of the helically driven system the helicity ratio for the small scale $E_M$ remains high in the early time regime. The ratio begins to drop as that of the large scale $E_M$ elevates. In contrast for the non-helical forcing, the helicity ratio of small scale $E_M$ remains low. And the helicity ratio of the large scale magnetic energy keeps growing after the initial fluctuation from the negative to the positive. Fig.~\ref{f11} shows the ratio of the large scale magnetic energy between the nonhelically driven and helically driven system, i.e., $E_{M, nonhelical}/E_{M, helical}(\equiv f_{ML})$ ($k=1$, red solid line) and the ratio of the residual helicity $\alpha_{nonhelical}/\alpha_{helical}$ on the small scale ($k=4-k_{max}$, black dashed line. Here we assumed $\tau_{helical}\sim \tau_{nonhelical}$). The relative ratio of the residual helicity remains lower than 0.1 but $f_{ML}$ is neither low nor constant. For $t<\sim100$ $E_{M, nonhelical}$ is larger than $E_{M, helical}$ (up to $\sim$ 100 times) and continues growing in spite of the low $\alpha$ effect. This result definitely indicates that the inverse transfer of $E_{M, nonhelical}$ is initially not related to the $\alpha$ effect. Rather it is related to other process like the local transfer through $-\textbf{u}\cdot\nabla \textbf{b}$. However, $f_{ML}$ keeps decreasing. And the $\alpha$ effect catches up with the effect of local energy transfer by $t\sim100$ so that $E_{M, helical}$ and $E_{M, nonhelical}$ become equal in their magnitudes. For $t> \sim100$, $E_{M, helical}$ surpasses $E_{M, nonhelical}$, which means the $\alpha$ effect becomes more efficient eventually. The helicity ratios of large scale $E_{M, nonhelical}$ and $E_{M, helical}$ at $t\sim100$ are in the range of $\sim0.45-0.5$.

\section{Theoretical Analysis}
The aim of this section is to investigate the inverse transfer of decaying (non)helical energy theoretically. We briefly introduce the statistical approach and scaling invariant method that tell us the fundamental principles of energy distribution in the MHD system. And then, we use the $\alpha$ dynamo theory to reproduce and explain the numerical results of the decaying helical magnetic field in the large scale regime. For the decaying nonhelical case, we modify and solve the Eddy Damped Quasi Normalized Makovianized approximation in a limited way \citep{Kraichnan and Nagarajan 1967, Pouquet et al 1976}. In addition to these analytic methods, we suggest a field structure model to explain the general mechanism of magnetic energy transport.\\

\subsection{Statistical approach and Scaling Invariant method}
The conserved physical quantities in a system decide the statistically most probable state of the system. This principle is also valid in the MHD turbulent system so that the most probable state of energy distribution is related to the properties of conserved system variables like energy, vector potential, or magnetic (cross) helicity. Lagrangian-Hamiltonian principle indicates that the conservation of a physical quantity is basically the result of symmetry in nature \citep{Goldstein et al 2013}. However, the MHD turbulence (e.g., dynamo) shows some contrariety to this generally accepted view. In the non cartesian coordinate system, the self consistent dynamo process between the poloidal and toroidal field is impossible with the axisymmetry \citep{Cowling 1934}. And the large scale dynamo with helicity cannot occur with the reflection symmetry \citep{Moffatt 1980, Krause and Radler 1980}. Interestingly the scaling invariant MHD equation \citep{Olesen 1997} shows that the distribution of energy is the intrinsic property of MHD equation regardless of the subtle distinction between Lagrangian-Hamiltonian principle and MHD equation. Now considering a simple example will be useful to have a holistic view of the basic principle in the statistical approach.\\

\noindent In a (statistically) closed system, the ensemble theory describes the stationary system that has the conserved variable $X_i$. The mean value of system variable $\langle X_i\rangle$ can be derived from Gibbs partition function `$Z=Z_0\, exp\,(-\sum \lambda_iX_i)$ ($i=1,\,2,\,3,...,\,N$)', where `$\lambda_i$' is a Lagrange multiplier corresponding to $\langle X_i\rangle$ (Frisch et al. 1975, Fyfe and Montgomery 1976, and references therein). In fact the relation between $\lambda_i$ gives us the information on the variable and system. In a 2D MHD system, the energy `$\sum_k \langle \omega_k^2+j_k^2\rangle/2k^2(\rightarrow \lambda_1$) ($\omega_k=\nabla \times \mathbf{u}_k,\,\mathbf{j}_k=\nabla \times \mathbf{b}_k)$', cross helicity `$\overline{P}\equiv \langle \omega_k\cdot \mathbf{j}_k\rangle/2k^2(\rightarrow \lambda_2)$', and vector potential `$\overline{A}\equiv \langle j_k^2\rangle/2k^4(\rightarrow \lambda_3)$' are conserved. With these conserved system variables we can construct the Gibbs distribution function like $Z=Z_0\, exp[-\lambda_1\overline{\epsilon}-\lambda_2\overline{P}-\lambda_3\overline{A}]$, which gives us information on the system. For example, the average of magnetic and kinetic energy from this Gibbs function are
\begin{eqnarray}
\langle b_k^2\rangle&=& \big(\lambda_1+\frac{\lambda_3}{k^2}-\frac{\lambda_2^2}{4\lambda_1}\big)^{-1},\label{EM 2D}\\
\langle u_k^2\rangle &=& \frac{1}{\lambda_1}+\frac{\lambda_2^2}{4\lambda_1^2}
\big(\lambda_1+\frac{\lambda_3}{k^2}-\frac{\lambda_2^2}{4\lambda_1}\big)^{-1}.
\label{EV_2D}
\end{eqnarray}
Eq.~(\ref{EM 2D}) shows that $\langle b_k^2\rangle$ has a peak at the minimum `$k$' if $E_M>E_V$ without the cross helicity ($\lambda_1>0$, $\lambda_2=0$, and $\lambda_3/\lambda_1<0$), which implies the inverse cascade of $E_M$. In contrast if $E_M < E_V$ ($\lambda_3/\lambda_1>0$), $E_M$ has the peak at the maximum $k$, or forward cascade of $E_M$. For the 3D MHD system the vector potential is replaced by the magnetic helicity $H_M(=\langle \mathbf{A}\cdot \mathbf{B}\rangle)$. But the basic principle is the same.\\

%
%
%

\noindent \cite{Olesen 1997} noted that the MHD equation itself is invariant with the scaled variables. The author used the property of the MHD equations to be invariant with respect to scaled variables (scaling invariant method): $\mathbf{r}\rightarrow l\mathbf{r}$, $t\rightarrow l^{1-h}t$, $\mathbf{u}\rightarrow l^h\mathbf{u}$, $\nu\rightarrow l^{1+h}\nu$, $\mathbf{b}\rightarrow l^h\mathbf{b}$, $\eta\rightarrow l^{1+h}\eta$, $P\rightarrow l^{2h}P$, where `$l$' and `$h$' are arbitrary parameters. Then the scaled kinetic energy is represented like
\begin{eqnarray}
&&\mathbb{E}_V(k/l, l^{1-h}t, Ll, K/l)\nonumber\\
&&=l^4 \frac{2\pi k^2}{(2\pi)^3}\int^L_{2\pi/K}d^3xd^3y\,\,e^{i\mathbf{k}\cdot (\mathbf{x}-\mathbf{y})}\langle \mathbf{u}(l\mathbf{x}, l^{1-h}t)\mathbf{u}(l\mathbf{y}, l^{1-h}t)\rangle\nonumber\\
&&= l^{4+2h}\mathbb{E}_V(k,t,L,K).\label{decaying E_V}
\label{decaying E_M}
\end{eqnarray}
Similarly the scaled magnetic energy can be found `$\mathbb{E}_M(k/l, l^{1-h}t, Ll, K/l)=l^{4+2h}\mathbb{E}_M(k,t,L,K)$'. Using a unknown function `$\psi (k,\,t)$' Olesen alternatively represented these energy density expressions `$E_{V,\,M}(k,\,t)=k^{-1-2h}\psi_{V,\,M} (k,\,t)$' leading to the relation of `$\psi_{V,\,M}(k/l,\,l^{1-h}t)=\psi_{V,\,M} (k,\,t).$' If these equations are differentiated with respect to `$l$' and then set `$l=1$', a differential equation is derived like
\begin{eqnarray}
-k\frac{\partial \psi_{V,\,M}}{\partial k}+(1-h)t\frac{\partial \psi_{V,\,M}}{\partial t}=0.
\label{psi differential equation}
\end{eqnarray}
The general solution of this equation shows `$\psi_{V,\,M} (k,\,t)$' is the function of `$k^{1-h}t$', which implies the (inverse) transfer of energy with increasing time `$k\sim t^{1/(h-1)}$'. If the initial energy spectrum is `$E_{V,\,M} (k,\,0)=k^{-1-2h}$', the energy at `$t$' is `$E_{V,\,M} (k,\,t)=k^{-1-2h}\psi_{V,\,M}(k^{1-h}t)=k^{q}\psi_{V,\,M}(k^{(3+q)/2}t),\,(q\equiv-1-2h)$'. This relation shows that when $h<1$ (i.e., $q>-3$), the decaying energy can be transferred inversely. However regardless of the general criterion for the migration of energy, the evolution of $E_M$ and $E_V$ are basically same. In fact 
the energy relation is not valid in the whole range. As \citep{Ditlevsen et al. 2004} pointed out the integration of $E_{V,\,M}\sim k^q\,\psi_{V,\,M}(k^{(3+q)/2}t)$ to get the total energy yields an inconsistent result. Moreover it is not yet clear if the scaled energy density relation can be applied to the helical field MHD system. Recently \cite{Brandenburg and Kahniashvili 2017} numerically tested the inverse transfer of energy in HD and MHD (helical and nonhelical) case with large and small $Pr_M$. Their result shows two types of initial energy distribution $E_M\sim k^4$ \& $E_V\sim k^2$ showing the tendency of inverse transfer, which was also found by \cite{Kahniashvili et al 2013}.

\begin{figure*}
  {\centering
   \includegraphics[width=8cm]{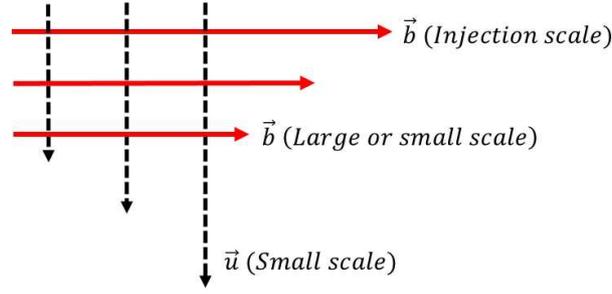}
   \caption{Coupled kinetic and magnetic fields. The red solid arrow lines indicate $E_M$, and the black dashed arrow ones indicate $E_V$. The length of arrow indicates the energy strength, not the eddy scale. The longest red solid line indicates the initial largest $E_M$ given to the magnetic eddy ($k=5$).}
   \label{f12}
  }
\end{figure*}

\begin{figure*}
\centering{
     \includegraphics[width=14cm]{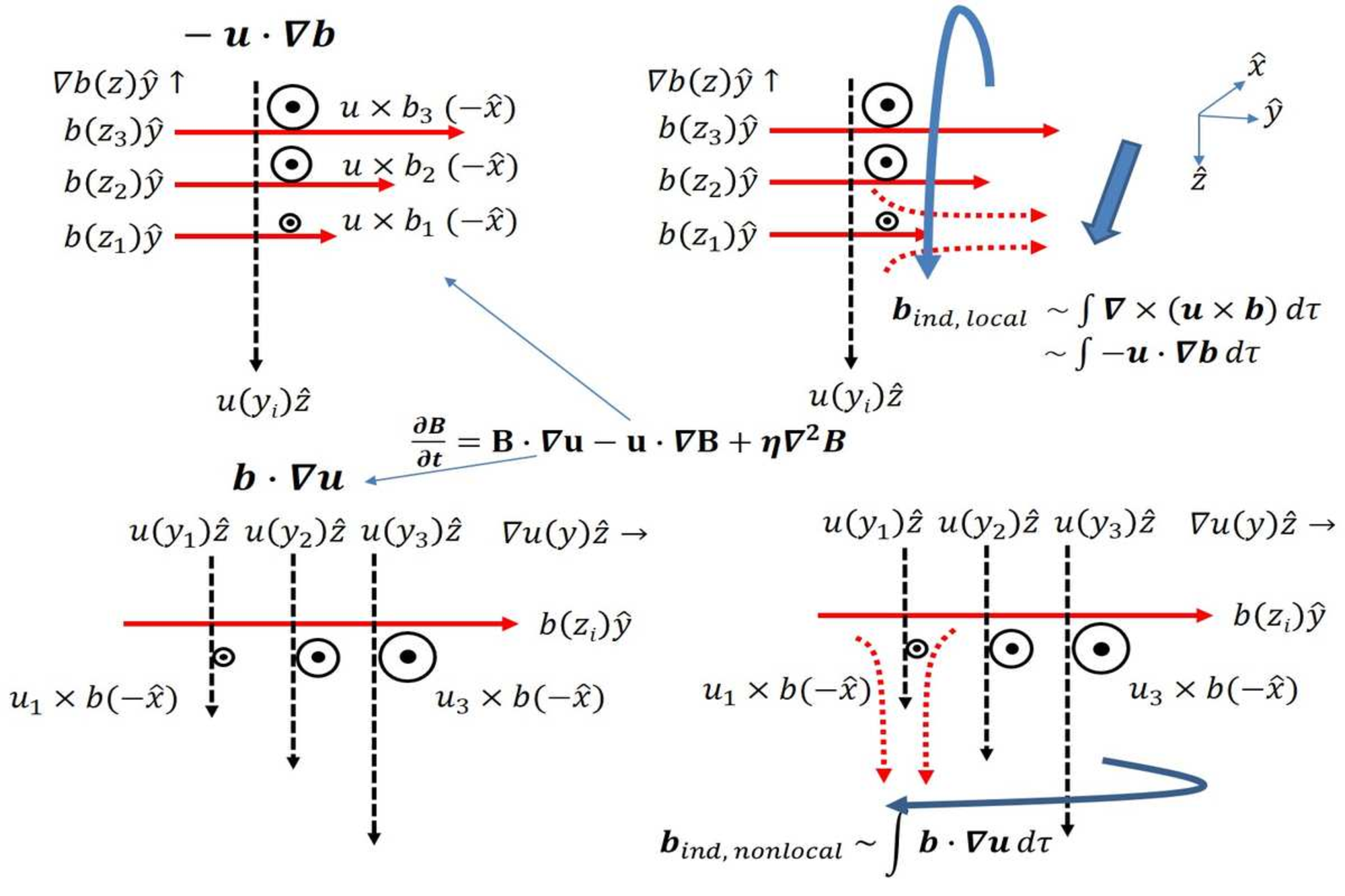}
     \label{f13}
     \caption{Field structure in the upper panel shows how the energy is transferred through $-\mathbf{u}\cdot \nabla \mathbf{b}$, and the lower panel shows how the energy is transferred through $\mathbf{b}\cdot \nabla \mathbf{u}$.}
     }
\end{figure*}

\begin{figure*}
\centering{
     \includegraphics[width=9cm]{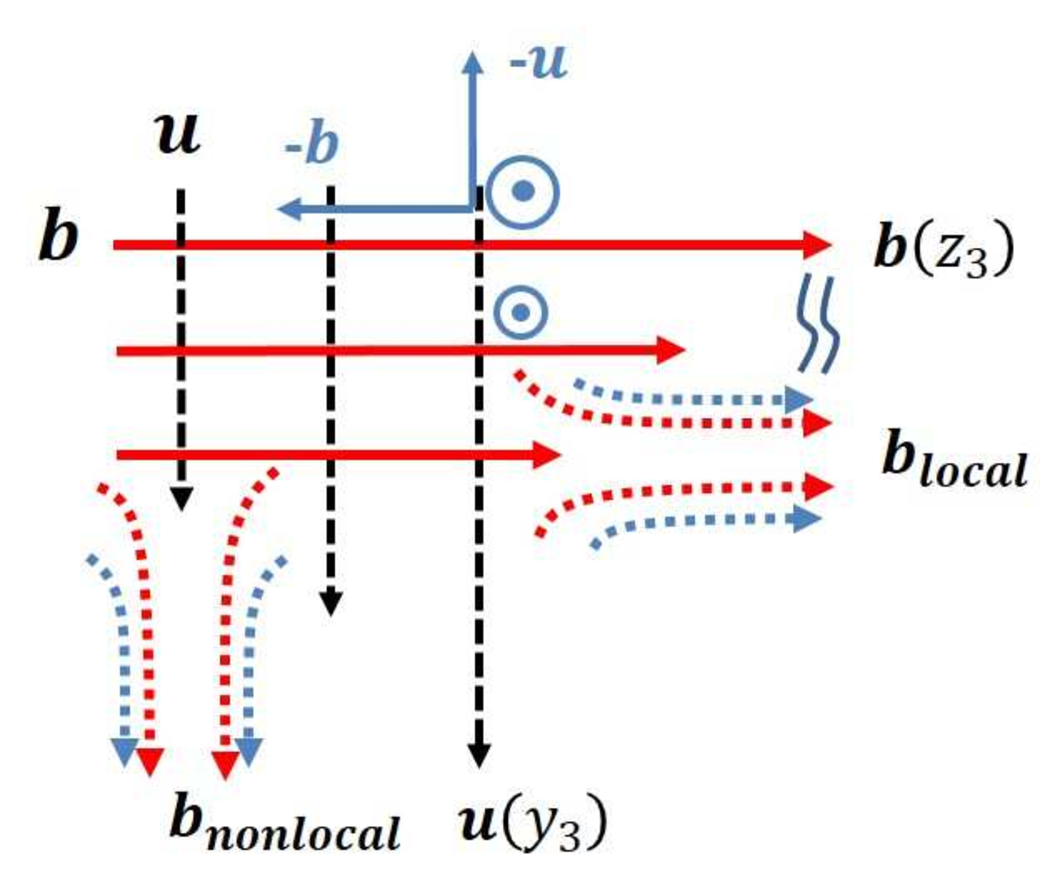}
     \label{f14}
\caption{The interaction of $-\mathbf{u}$ and $-\mathbf{b}$ in the small scale ($z_3$) induces the additional EMF at $\mathbf{b}(z_i)$ and $\mathbf{u}(z_i)$ (i=1, 2).}}
\end{figure*}

\subsection{Inverse Transfer of Decaying Helical Magnetic Energy}
The simulation results of inverse cascade for the decaying helical magnetic field was initially reported by \citet{Biskamp and Mueller 1999, Christensson et al 2001}. Their motivation was to explain how the primordial magnetic fields could be amplified into large and strong magnetic fields for seeding the galactic dynamo. Their numerical results showed the inverse cascade of decaying helical magnetic energy in Fourier space. However the qualitative analysis used in their work is not enough to describe or analyze the inversely cascading magnetic energy. So in this section we reproduce the evolving large scale magnetic field using the semi-analytic $\alpha^2$ dynamo theory for the comparison with the numerical results.\\

\noindent For a helical  MHD system, $\alpha$ dynamo theory is advantageous for its flexible application to the (non)ideal helical MHD system. With the helicity in the small scale regime, the large scale magnetic field $\overline{\mathbf{B}}~(=\mathbf{B}-\mathbf{b}$, $\mathbf{b}$: small scale magnetic field) can be represented as  \citep{Krause and Radler 1980, Moffatt 1980}
\begin{eqnarray}
\frac{\partial {\bf \overline{B}}}{\partial t}&=&\nabla \times \langle {\bf u}\times {\bf b}\rangle + \eta \nabla^2{\bf \overline{B}},\\
\label{Exact_Mean large scale magnetic field}
&\sim&\nabla \times \alpha {\bf \overline{B}} + (\eta+\beta) \nabla^2{\bf \overline{B}}.
\label{Approximate_Mean large scale magnetic field}
\end{eqnarray}
where $\alpha=1/3\int^t_{-\infty}\big(\langle {\bf j}\cdot {\bf b}\rangle-\langle {\bf u}\cdot {\bf\omega}\rangle\big ) dt$ and $\beta=1/3\int^t_{-\infty}\langle u^2\rangle\,dt$. The helical kinetic or magnetic energy in small scale regime, i.e., $\alpha$ interacts with $\overline{\mathbf{B}}$ leading to the induction (evolution) of $\overline{\mathbf{B}}$. In contrast $\beta$ from the kinetic energy on small scales plays the role of a dissipation coefficient. Abut at the same time the plasma motion ($\sim u$) is a prerequisite condition of the induction of magnetic field.\\

\noindent Eq.~(\ref{Approximate_Mean large scale magnetic field}) implies that the large scale magnetic field can grow regardless of the forcing source as long as the $\alpha$ effect is stronger than that of magnetic dissipation. However, this kind of conceptual understanding is theoretically incomplete and practically not so much useful. As the equation indicates, the current density $\overline{\bf J}(=\nabla \times \overline{\bf B})$ is the source of magnetic field $\overline{\bf B}$ so that the coupled equation of large scale $\overline{E}_M$ and $\overline{H}_M$ $(\equiv \langle \overline{\mathbf{A}}\cdot \overline{\mathbf{B}}\rangle$, more exactly $k^2\overline{H}_M$) are derived due to the different derivative order in $\overline{\bf B}$ and $\overline{\bf J}$. And the coupling strength due to the relative difference between $\overline{E}_M$ and $\overline{H}_M$ changes with time affecting the saturated $\overline{E}_M$ and $\overline{H}_M$. So instead of the qualitative analysis for Eq.~(\ref{Approximate_Mean large scale magnetic field}), the coupled $\overline{E}_M$ and $\overline{H}_M$ should be solved simultaneously.\\

\noindent From Eq.~(\ref{Approximate_Mean large scale magnetic field}) $\overline{E}_M (\equiv\langle\overline{\mathbf{B}}\cdot \overline{\mathbf{B}} \rangle/2)$ and $\overline{H}_M (\equiv\langle\overline{\mathbf{A}}\cdot \overline{\mathbf{B}})$ are derived like
\begin{eqnarray}
\frac{\partial }{\partial t}\overline{E}_M
&=&\alpha\overline{H}_M-2(\beta+\eta)\overline{E}_M,\label{Em with alpha coefficient}\\
\frac{\partial }{\partial t}\overline{H}_M
&=&4\alpha \overline{E}_M-2(\beta+\eta)\overline{H}_M. \label{Hm with alpha coefficient}
\end{eqnarray}
Through the transformation of bases we can find the solutions like below:
\begin{eqnarray}
2\overline{H}_M(t)&=&(\overline{H}_M(0)+2\overline{E}_M(0))e^{2\int^t_0(\alpha-\beta-\eta)dt'}\nonumber\\
&&+(\overline{H}_M(0)-2\overline{E}_M(0))e^{-2\int^t_0(\alpha+\beta+\eta)dt'},\label{HmSolution1}\\
4\overline{E}_M(t)&=&(\overline{H}_M(0)+2\overline{E}_M(0))e^{2\int^t_0(\alpha-\beta-\eta)dt'}\nonumber\\
&&-(\overline{H}_M(0)-2\overline{E}_M(0))e^{-2\int^t_0(\alpha+\beta+\eta)dt'}.\label{EmSolution2}
\end{eqnarray}
(Here $\overline{E}_M(0)$ and $\overline{H}_M(0)$ are the measured large scale magnetic energy and helicity at $t=1$ when the forcing is turned off.)\\

\noindent Since the system was initially driven by the positive magnetic helicity ($\alpha>0$), the influence of the second term is ignorably small. Thus the large scale magnetic energy and helicity converge to\footnote{When the analytic equation was plotted, the time interval $\Delta t=t_{n}-t_{n-1}$ and injection scale range $k=4-10$ were considered for the practical reason.}

\begin{eqnarray}
\overline{H}_M(t)&=&\frac{1}{2}(\overline{H}_M(0)+2\overline{E}_M(0))e^{2\int^t_0(\alpha-\beta-\eta)dt'}
,\label{HmSolution1a}\\
\overline{E}_M(t)&=&\frac{1}{4}(\overline{H}_M(0)+2\overline{E}_M(0))e^{2\int^t_0(\alpha-\beta-\eta)dt'}
.\label{EmSolution2a}
\end{eqnarray}
In contrast, if the given initial magnetic helicity is negative, the second term becomes dominant. The evolution of field profile and other features are the same as those of the previous case except the opposite sign of the magnetic helicity. Now Eq.~(\ref{HmSolution1a}), (\ref{EmSolution2a}) can be compared with the simulation data to check its validness. To do so, we used the numerical data in the small scale regime for the $\alpha$ \& $\beta$ coefficient. The semi-analytic equation is included in Fig.~\ref{f2} for the comparison with the numerical result. The analytic equation and numerical result match well eventually. But there is some discrepancy in the early time regime when the $\alpha$ effect ($\sim \alpha \overline{\mathbf{B}}$) is not yet strong enough. This means that $E_M$ in the early time regime is transferred through a different principle rather than the $\alpha$ effect.\\

\noindent On the other hand, Fig.~\ref{f2} also shows the evolution of nonhelical large scale magnetic energy $\overline{E}_M-\langle \overline{\mathbf{A}}\cdot \overline{\mathbf{B}}\rangle/2$, which can be derived and represented like below:
\begin{eqnarray}
\overline{E}_{M,\,NH}(t)&=&\frac{1}{2}\big(2\overline{E}_M(0)-|\overline{H}_M(0)|\big)
e^{-2\int^t_0(|\alpha|+\beta+\eta)dt'}.
\label{Nonhelical_Em}
\end{eqnarray}
This solution shows that the nonhelical $\overline{E}_M$ disappears exponentially due to the $\alpha$ effect and plasma motion. Here we used the simulation data for $\alpha$ \& $\beta$ coefficient. But these coefficients can be obtained from the observed data.

\subsection{Inverse Transfer of Decaying Nonhelical Magnetic Energy}
We have introduced the ensemble theory and scaling invariant method to explain the mechanism of energy transfer conceptually. However, they are not suitable to explain the evolving profile of decaying nonhelical magnetic field. Therefore we first suggest a field structure model based on the magnetic induction equation for the intuitive understanding $E_M$ transfer. And then we solve the quasi normal approximation for the bi-directional transfer of magnetic energy across different spatial scales.

\subsubsection{Field Structure}
\noindent Before discussing the model we recall in the following a few fundamental points. First, the conventional MHD theory describes the MHD system with the fluid-dynamical point of view and concepts. It may give some easy picture of the energy transfer. However, the magnetic field does not flow like liquid. Strictly speaking, the magnetic field is induced through the interaction between plasma motion and seed magnetic field. So the usage of hydrodynamic concepts for the magnetic field requires careful attention. Second, the dynamo theory usually assumes that the system is statistically homogeneous and isotropic for simplicity. However, if the magnetic and velocity field are completely homogeneous and isotropic, the dynamo process cannot occur. Indeed the velocity and magnetic field should be at least locally inhomogeneous and anisotropic for the nontrivial dynamo process. Third, we do not consider the effect of cross helicity except for some conceptual explanations in this paper.\\

\noindent In Fig.~5 we construct the `field structure' based on the mathematical definition of $\nabla\times\langle \mathbf{u}\times \mathbf{b} \rangle\sim - \mathbf{u}\cdot \nabla \mathbf{b} + \mathbf{b}\cdot \nabla \mathbf{u}$ (in an approximately incompressible system). The first term `$-\mathbf{u}\cdot \nabla \mathbf{b}$' means that the velocity field heads toward the decreasing magnetic field, and `$\mathbf{b}\cdot \nabla \mathbf{u}$' means that the magnetic field heads toward the increasing velocity field. The actual field behaviors are even more complicated, but this simple geometry induces the magnetic field most efficiently. And the length of each arrow in the figure indicates the strength of the energy ($\sim B,\,u$), not the eddy scale. Yokoi \citep{Yokoi 2013} used a similar structure to explain the generation of current density. But our interest in this paper is not the current density but the migration (induction) of magnetic field. The generation of helical current density will be derived for the $\alpha$ effect and conservation of magnetic helicity in the subsequent paper, not in this paper.\\

\noindent The upper panel in Fig.~6 shows how the energy is transferred through the local transfer term $-\mathbf{u}\cdot \nabla \mathbf{b}$. The magnetic field $\mathbf{b}(z_i)$ (i=1, 2, 3) heads toward the $\hat{y}$-direction, and is a function of $z_i$ with $|b(z_1)| < |b(z_2)| < |b(z_3)|$, where $z_1 > z_2 > z_3$. As mentioned this field structure itself does not impose any constraint on the eddy scale, so $b(z_3)$ could be considered as a large or small scale magnetic field without any loss of generality. However since we have given the initial energy to the injection scale ($k=5$, small scale regime), we consider $b(z_3)$ as a small scale magnetic eddy. We refer to $b(z_1)$ as a large scale magnetic eddy ($k=1)$ for the inverse transfer. But like $b(z_3)$ we can also consider $b(z_1)$ as a smaller scale magnetic eddy for the forward transfer. The point is $b(z_i)$ in this model is proportional to $\sim E_M^{1/2}(z_i)$, not the eddy scale.\\

\noindent We assume a simple structure of the velocity field $\mathbf{u}(y_i)=(0,\,0,\,u)$ and the magnetic field $\mathbf{b}(z_i)=(0, \,b(z_i), \,0)$ where $\partial b(z_i)/\partial z\,<\,0$. This local transfer term $-\mathbf{u}\cdot \nabla \mathbf{b}$ is simply represented like
\begin{eqnarray}
-u\,\hat{z}\cdot \bigg(\hat{x}\frac{\partial}{\partial x}+\hat{y}\frac{\partial}{\partial y}+ \hat{z}\frac{\partial}{\partial z}\bigg)b(z_i)\,\hat{y}=-u\,\frac{\partial b(z_i)}{\partial z}\,\hat{y}.
\label{u_cdot_nabla_b}
\end{eqnarray}
This result $-u\,\partial b(z_i)/\partial z>0$ indicates that the magnetic energy is induced at $z_i$ and sequentially transferred to its neighboring eddy: $b(z_3)\rightarrow b(z_2) \rightarrow b(z_1)$. The direction of induced magnetic field is parallel to $\mathbf{b}(z_i)$, and direction of magnetic energy transfer is perpendicular to $\mathbf{b}(z_i)$. This process can be explained more clearly with the concept of the curl operator. The interaction between $\mathbf{u}$ and $\mathbf{b}$ generates an EMF varying from the strongest $\langle u\times b(z_3) \rangle\, (-\hat{x})$ to the weakest $\langle u\times b(z_1)\rangle\, (-\hat{x})$. Then their differential strengths create a non-trivial curl effect, which induces a magnetic field: $\mathbf{b}_{ind,\,local}\sim \int d\tau\, \nabla \times \langle \mathbf{u}\times \mathbf{b}(z_i)\rangle$ (i=1, 2) (Fig.~6).\\

\noindent In contrast, if $\mathbf{b}(z_1)$ is a small scale magnetic field, the energy transfer indicates forward cascade of $E_M$. The direction of local magnetic energy transport is not limited to one direction, forward or inverse. The bidirectional transfer of magnetic energy is clearly shown by evolving $E_M$ of $k=1,\,8$ in Fig.~\ref{f2}, \ref{f6}.\\

\noindent This structure also implies a constraint on the energy transport. As the induced magnetic field at $\mathbf{b}(z_1)$ grows, the curl effect decreases so that less energy is transferred toward $\mathbf{b}(z_1)$. In principle, the magnetic energy transport is limited by $b(z_3) \gtrsim b(z_1)+b_{ind,\,l}$. However if the field is helical, this constraint is not valid. The growth rate is proportional to the strength of net magnetic field at $z_1$, the transfer of $E_M$ from $\mathbf{b}(z_3)$ is accelerated as the magnetic field at $z_1$ grows. This explains the increasing growth rate of the helical $E_{M,\,L}$ with time (Fig.~\ref{f11}).\\

\noindent The lower part in Fig.~6 shows how the magnetic energy is transferred through the non-local transfer term $\mathbf{b}\cdot \nabla \mathbf{u}$. Similarly we define the magnetic field and velocity field as $\mathbf{b}=(0,\,b,\,0)$ and $\mathbf{u}(y_i)=(0, \,0,\,u(y_i))$. The strength of the velocity field is assumed to be in the order of $u(y_1) < u(y_2) < u(y_3)$ with $y_1 < y_2 < y_3$. Here, $u(y_3)$ is assumed to be a small scale velocity field with an initial large energy. This nonlocal term is represented as follows:
\begin{eqnarray}
\mathbf{b}\cdot \nabla \mathbf{u}\rightarrow b\,\hat{y}\cdot \bigg(\hat{x}\frac{\partial}{\partial x}+\hat{y}\frac{\partial}{\partial y} + \hat{z}\frac{\partial}{\partial z}\bigg)u(y_i)\,\hat{z}=b\,\frac{\partial u(y_i)}{\partial y}\,\hat{z},
\label{b_cdot_nabla_u}
\end{eqnarray}
which is positive. Then, $\mathbf{b}\cdot \nabla \mathbf{u}$ induces the magnetic field $\mathbf{b}_{ind,\,nl}$. The direction of $\mathbf{b}_{ind,\,nl}$ is parallel to $\mathbf{u}(y_i)$, and the direction of $E_M$ transport is parallel to $\mathbf{b}(z_i)$. We can also more clearly explain the cascade of $E_M$ using the cross product and curl effect. The EMF varies from $\langle \mathbf{u}(y_1)\times \mathbf{b}\rangle$ to $\langle \mathbf{u}(y_3)\times \mathbf{b}\rangle$ in the increasing order of magnitude, whose direction is out of the page. Then $\nabla \times \langle \mathbf{u}(y_i)\times \mathbf{b}\rangle$ is the strongest at $\mathbf{u}(y_1)$. The strength of net magnetic field induced by this local and nonlocal transfer term is the vector summation of these induced fields: $b_{net}=\sqrt{b^2_{ind,\,l}+b^2_{ind,\,nl}}$. The scale of this net magnetic field can be larger (inverse transfer) or smaller (forward transfer) than that of the seed $u$, $b$ as mentioned.\\

\begin{figure*}
\centering{
  {
   \subfigure[Inverse Cascade I]{
     \includegraphics[width=3.6cm]{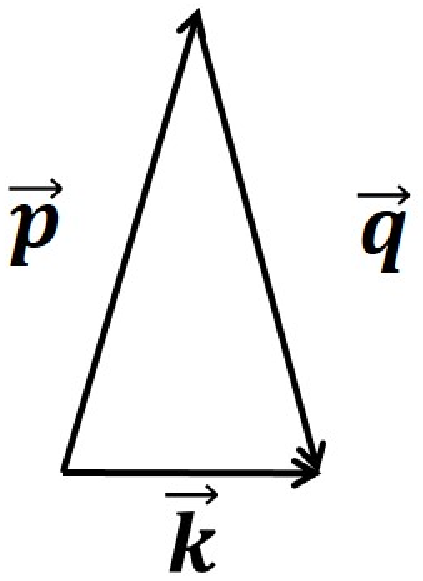}
     \label{triangle1}
   }\hspace{-2mm}
   \subfigure[Inverse Cascade II]{
     \includegraphics[width=6.6cm]{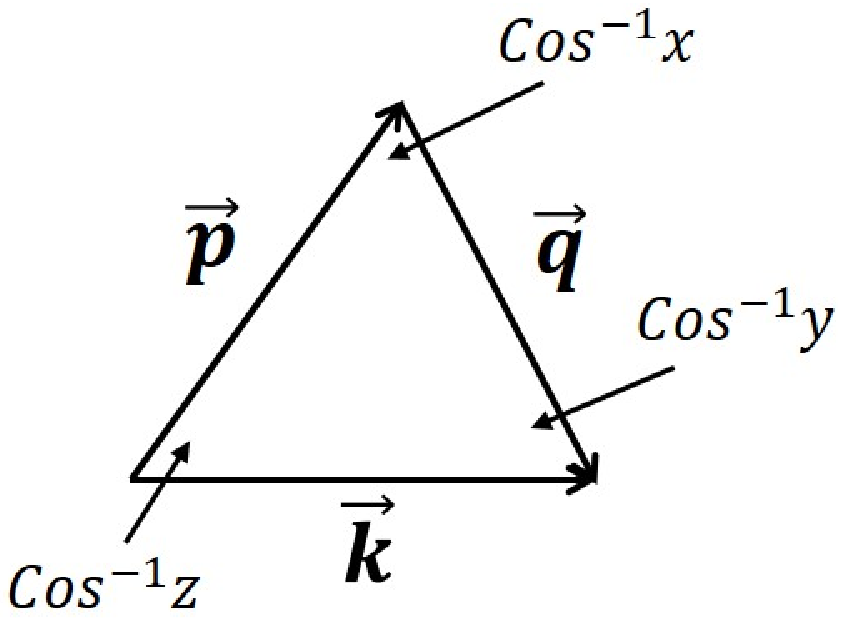}
     \label{triangle2}
   }\hspace{0mm}
   \subfigure[Forward Cascade]{
     \includegraphics[width=6cm]{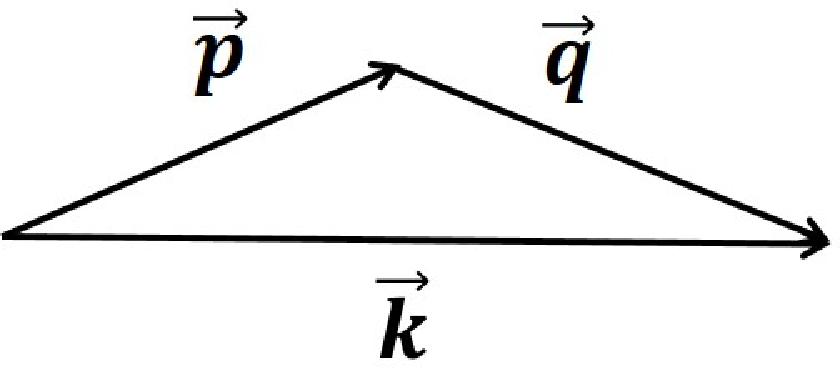}
     \label{triangle3}
     }
  }
\caption{}
}
\end{figure*}


%
\subsubsection{Role of $\tau$ in Energy Transport}
So far we have shown that $E_M$ can migrate toward the larger and smaller scale. But there are two more things that explain the inverse transfer of $E_M$ in a decaying MHD system. First, since the small scale eddies ($\mathbf{u}(y_3)$ and $\mathbf{b}(z_3)$) have large energy, the dissipation effect is also large causing the fast decay. As Fig.~7 shows, these rapidly decaying fields can be considered as a growth of the velocity or magnetic field in the opposite direction: $-u(y_3)\,\hat{z}$ or $-b(z_3)\,\hat{y}$ (blue solid line). Then the additional curl effect arises and boosts the induction of magnetic field at $\mathbf{u}(y_1)$ and $\mathbf{b}(z_1)$. In fact, the fast decaying $\mathbf{u}(y_3)$ and $\textbf{b}(z_3)$ substantially increase $\Delta \mathbf{u}(y)$ and $\Delta \mathbf{b}(z)$ in Eq.~(\ref{u_cdot_nabla_b}), (\ref{b_cdot_nabla_u}), which increases the curl effect. In contrast if $u(y_3)$ or $b(z_3)$ is a large scale field, the effect due to the fast decay cannot be expected much.\\

\noindent Also `$\tau$' plays an important role in the migration of $E_M$. From the Fourier transformed magnetic induction equation
\begin{eqnarray}
\frac{\partial \mathbf{b}(\mathbf{k},\,t)}{\partial t}=\sum_{\mathbf{p},\,\mathbf{q}}\nabla \times \big\langle \mathbf{u}(\mathbf{p},\,t)\times \mathbf{b}(\mathbf{q},\,t)\big\rangle\,\delta(\mathbf{p}+\mathbf{q}-\mathbf{k})-\eta k^2 \mathbf{b}(\mathbf{k},\,t),
\label{Curl of EMF in Fourier space}
\end{eqnarray}
we may simply infer that the magnetic field $\mathbf{b}(\mathbf{k})$ is induced through the interaction between $\mathbf{u}(\mathbf{p})$ and $\mathbf{b}(\mathbf{q})$ with the constraint of $\mathbf{k}=\mathbf{p}+\mathbf{q}$. However, the magnetic induction equation $\partial \mathbf{b}/\partial t\sim \nabla \times \langle\mathbf{u}\times \mathbf{b}\rangle \sim -\mathbf{u}\cdot \nabla \mathbf{b}+\mathbf{b}\cdot \nabla \mathbf{u}$ implies that `$-1/\mathbf{u}\cdot\nabla$' has the dimension of time `$\tau_u$.' Also the dimension of the Alvf$\acute{e}$n time $\tau_A=L/v_A\,(v_A\sim B)$ and the linearized Els\"{a}sser variable equation `$\partial_t\mathbf{z}^{\pm}-\mathbf{B}\cdot\nabla \mathbf{z}^{\pm}=0\, (\mathbf{z}^{\pm}=\mathbf{u}\pm \mathbf{b})$' imply that  `$\mathbf{b}\cdot \nabla$' has the dimension of time `$\tau_b$'\citep{Biskamp 2008, Montgomery and Turner 1981}. Then the solution of the magnetic induction equation is basically a function of superposed exponentials $\sim exp(\pm\, t/\tau_{u,\,b})$ ($\tau_u\sim 1/ku$ and $\tau_b\sim1/kb$). If $\tau_u$ or $\tau_{b}$ is large (small energy or small wave number), a longer time is required to reach the same amount of energy. Again the decreasing energy transport elevates `$\tau$' recursively. In contrast the growing energy decreases $\tau$ leading to the more efficient energy transport.\\

\noindent In the large scale regime the relatively large $\tau$ due to the small wavenumber resists the change of energy  initially. But because of the low dissipation effect, energy does not decay so fast that $\tau$ does not grow fast. Moreover the rate of energy transfer from the magnetic eddy to the plasma on the large scale regime decreases as the magnetic helicity ratio grows naturally (Fig.~\ref{f10}, Eq.~(\ref{Gen of A.B1}), (\ref{Gen of A.B2})). As a result the increase of large scale energy through the inverse transfer accelerates.\\

\noindent The comprehension of the energy cascade only within the framework of $\tau$ is too limited. Since the magnetic energy transfer is basically the result of the interaction between $E_V$ and $E_M$, a more detailed investigation of the effect of $E_V$ on the energy transfer is required. At this moment we will just check the tendency shown in the simulation data. Figs.~\ref{Tauhelical}, \ref{TauNonhelical} show that the evolution of $\tau_u\sim 1/|\mathbf{u}\cdot \nabla|$ precedes that of $\tau_b\sim 1/|\mathbf{b}\cdot \nabla|$. For example, in case of the decaying helical field (Fig.~\ref{Tauhelical}) $\tau_u$ at $k=7$ (decreasing) and $k=4$ (decreasing) meet each other and change their order at $t\sim1-2$: $\tau_u(k=7)<\tau_u(k=4)\,\rightarrow \, \tau_u(k=7)>\tau_u(k=4)$. And at $t\sim10$ the similar turning around of $\tau_b$ and $E_M$ occurs: $\tau_b(k=7)<\tau_b(k=4)\,\rightarrow \, \tau_b(k=7)>\tau_b(k=4)$, $E_M(k=7) > E_M(k=4)\,\rightarrow \, E_M(k=7) < E_M(k=4)$. And at $t\sim3-4$ we see $\tau_u(k=7)<\tau_u(k=1)\,\rightarrow \, \tau_u(k=7)>\tau_u(k=1)$, which is followed by $\tau_b(k=7)<\tau_b(k=1)\,\rightarrow \, \tau_b(k=7)>\tau_b(k=1)$, $E_M(k=7) > E_M(k=1)\,\rightarrow \, E_M(k=7) < E_M(k=1)$ later. Fig.~\ref{TauNonhelical} shows that the trend in the decaying nonhelical system is in fact the same. These results indicate that the plasma motion leads the evolution of $E_M$ profile whether or not $E_V$ in the system is weak. However since the induced magnetic field `$\mathbf{b}$' is the vector summation of `$\mathbf{b}_{nonlocal}$' and `$\mathbf{b}(z_1)+\mathbf{b}_{local}$', the saturation of energy transfer is completed according to max($\tau_b$, $\tau_u$). Next we will discuss about the role of $E_V$ in the $E_M$ transfer analytically.\\


\subsubsection{Eddy Damped Quasi Normalized Markovian approximation}
The field structure model should be complemented by a more strict analytic method. However at present there is no self-consistent theory for the (inverse) transfer of nonhelical magnetic energy. nonetheless, the formal solution of EDQNM approximation gives us clues to the energy transfer in the MHD system. Since we are interested in the evolution of $E_M$ in a decaying MHD system with the infinity $Pr_M$, we consider only the equation of magnetic energy:
\begin{eqnarray}
\frac{\partial E_M(k,\,t)}{\partial t}&=&
\underbrace{-\int dp\,dq\,\Theta^{\eta\eta\nu}_{kqp}(t)\frac{p^2}{q}z(1-x^2)E_M(q,\,t)E_M(k,\,t)}_{(a)}\nonumber\\
&&\underbrace{-\int dp\,dq\,\Theta^{\eta\eta\nu}_{kqp}(t)\frac{q^2}{p}(y+xz)E_V(p,\,t)E_M(k,\,t)}_{(b)}\nonumber\\
&&\underbrace{+\int dp\,dq\,\Theta^{\eta\eta\nu}_{kqp}(t)\frac{k^3}{p\,q}(1+xyz)E_V(p,\,t)E_M(q,\,t)}_{(c)}.\nonumber\\
\label{EDQNMEm(t)}
\end{eqnarray}
The wave number $\mathbf{p},\,\mathbf{q},\,\mathbf{k}$ are constrained by $\mathbf{p}+\mathbf{q}=\mathbf{k}$, and $x,\,y,\,z$ are the cosines of angles formed by these three wave vectors (see Figs.~8). If $k<p,\,q$ with the positive right hand side (RHS) of Eq.~(\ref{EDQNMEm(t)}), the energy at $p$ and $q$ is inversely transferred to $E_M(k)$. In contrast, if $k>p,\,q$ and positive RHS, $E_M$ at $p$ and $q$ is forwardly transferred to $E_M(k)$.\\

\noindent Since EDQNM method uses an iterative method to close the equation, an eddy damping function $\Theta_{kqp}$ is necessary when the differentiated third order moment is integrated back. The eddy damping function is derived like
\begin{eqnarray}
\Theta_{kqp}^{\eta\eta\nu}(k,t)\sim\frac{1-e^{-\int^t\big(\nu k^2+\eta p^2+\eta q^2+\mu_{kpq}\big)\,dt'}}{\nu p^2+\eta(k^2+q^2)+\mu_{kpq}}.
\label{EddyDampingFunctionTriRelTime}
\end{eqnarray}
Here $\mu_{kpq}$, triad relaxation time, is sort of an error correction factor required when the fourth order moment $\langle vvvv\rangle$ is decomposed into the second order moments $\sum\langle vv\rangle\langle vv\rangle$. But we ignore $\mu_{kpq}$ here for simplicity. Moreover if we consider the system later than a few times of large scale $\tau$, $\Theta_{kqp}$ can be more simplified.\\

\noindent For magnetic energy in the very small scale regime to be transferred to the large scale eddy, one of the possible (nontrivial) relation of the wave vectors is shown in Fig.~\ref{triangle1} ($p$, $q \gg k$). Since the cosines of angles converge to $y\sim z\sim 0$, $x\sim 1$, `$(a)$', `$(b)$' in Eq.~(\ref{EDQNMEm(t)}) disappear. Only `$(c)$' has a nontrivial value:
\begin{eqnarray}
\frac{\partial E_M(k,\,t)}{\partial t}=\frac{k^3}{\nu p^3q}E_V(p,\,t)E_M(q,\,t)(1-e^{-\nu k^2t}).
\label{EDQNMInverse1}
\end{eqnarray}
This result shows the growth rate of magnetic energy at $k$ increases from $zero$ to a saturated value as $E_V(p,\,t)$ or $E_M(q,\,t)$ decreases. The energy is inversely transferred from $E_V(p)$ and $E_M(q)$ to $E_M(k)$ regardless of the helicity. In the nonideal MHD system, ($a$), ($b$) cannot be completely ignored. The initial growth rate becomes lower and converges to $zero$ faster.\\

\noindent For another example of inverse transfer, we can also consider the eddies whose scales are not so much different from that of $k=1$. With the condition of $p\sim q\gtrsim k$ ($y\sim z\sim x\rightarrow 1/2$, see Fig.~\ref{triangle2}), Eq.~(\ref{EDQNMEm(t)}) at $t\rightarrow \infty$ becomes
\begin{eqnarray}
\frac{\partial E_M(k,\,t)}{\partial t}&=&\underbrace{-\frac{1}{\nu}\bigg(\frac{3}{8q}E_M(q,\,t)+\frac{3q^2}{4p^3}E_V(p,\,t)\bigg)E_M(k,\,t)}
_{\equiv -P(p,\,q,\,t)E_M(k,\,t)}\nonumber\\
&&+\underbrace{\frac{1}{\nu}\frac{9k^3}{8p^3q}E_V(p,\,t)E_M(q,\,t)}_{\equiv Q(p,\,q,\,t)}.
\label{EDQNMInverse2}
\end{eqnarray}
The formal solution
\begin{eqnarray}
E_M(k,\,t)&=&e^{-\int^tPdt'}\big[\int^t e^{\int^{\tau}Pdt''} Qd\tau+E_M(k,\,0)\big]\nonumber\\
&\rightarrow& e^{-\int^tPdt'}\int^t e^{\int^{\tau}Pdt''} Qd\tau\nonumber\\
&\sim&\frac{1}{P}Q-\frac{1}{P^2}Qe^{-\int^tPdt'}\ldots
\label{EDQNMInverse2Solution}
\end{eqnarray}
shows that $E_M(k,\,t)$ grows due to $Q(p,\,q,\,t)$ in the smaller scale regime. And the growth of $E_M(k,\,t)$ is also affected by $P(p,\,q,\,t)$. In our case the initial $p,\, q$ are approximately 5, but $p,\,q$ change (spread) with time. The simulations shown in Fig.~2 are between these two extreme cases, Eq.~(\ref{EDQNMInverse1}), (\ref{EDQNMInverse2Solution}). These results also imply that the growing $E_M(k)$ suppresses the growth rate of $E_M(k)$.\\

\noindent On the other hand, the forward cascade of magnetic energy $p,\,q < k$ can be explained with different geometries. First, as Fig.~\ref{triangle3} shows, if $p\sim q < k\,(\rightarrow \infty)$, the angle relation $x\rightarrow -1$ \& $y,\,z\rightarrow 1$ makes the forward cascade impossible. But if $q\ll p \lesssim k$ or $p\ll q \lesssim k$, the forward cascade of $E_M$ occurs. The formal growth rate of $E_M(k,\,t)$ becomes similar to that of Eq.~(\ref{EDQNMInverse2}). These conditions imply the possibility of small scale dynamo. The largest wave number `$k$' from the interaction between the plasma motion and seed magnetic field is `$p+q$'. But in this case the electromotive force itself will be negligible, which does not guarantee any dynamo process.\\

\noindent For the exact solution, the coupled equations of $E_V(k,\,t)$ and $E_M(k,\,t)$ should be solved simultaneously. This can be done through the transformation of bases, which separates the coupled variables. The nontrivial secular equation will produce the relation between $E_V(k,\,t)$ and $E_M(k,\,t)$. But at present it is not clear whether the relation will produce a complete set or not. More detailed work is necessary. For now it is enough to show the energy at $p,\,q$ induces $E_M$ at $k=1$. Also we need to note that the term (c), source in Eq.~(\ref{EDQNMEm(t)}) cannot be negative. But (a), or (b) can be negative or positive according to the relation among $x$, $y$, and $z$ ($p$, $q$, $k$).

\begin{figure*}
\normalsize
\begin{eqnarray}
&&\frac{\partial b_{y}}{\partial t}=-u_{z}\frac{\partial b_y}{\partial z} + \eta\nabla^2 b_y\qquad \big(\rightarrow \frac{\partial \mathbf{b}_{pol}}{\partial t}=\underbrace{-\mathbf{u}_{tor}\cdot \nabla \mathbf{b}_{pol}}_{\rightarrow\, (\hat{\mathbf{r}},\,\hat{\mathbf{\theta}})\frac{\partial}{\partial\phi}(b_r,\,b_{\theta}),\,(\mathbf{1a})}+\eta\nabla^2\mathbf{b}_{pol}\big),
\label{poloidal field alpha-Omega dynamo}\\
&&\frac{\partial b_{z}}{\partial t}=+b_{y}\frac{\partial u_z}{\partial y} + \eta\nabla^2 b_z\qquad\big(\rightarrow
\frac{\partial \mathbf{b}_{tor}}{\partial t}=\underbrace{-\mathbf{u}_{tor}\cdot \nabla \mathbf{b}_{pol}}_{\rightarrow\,(b_r,\,b_{\theta})\frac{\partial}{\partial\phi}(\hat{\mathbf{r}},\,\hat{\mathbf{\theta}}),\,(\mathbf{1b})}+\underbrace{\mathbf{b}_{pol}\cdot \nabla \mathbf{u}_{tor}}_2+\eta\big(\nabla^2-\frac{1}{r^2\,\mathrm{sin}^2\theta}\big) \mathbf{b}_{tor}\big).
\label{toroidal field alpha-Omega dynamo}\\\nonumber\\
&&\qquad\qquad\qquad\qquad\qquad\qquad\qquad\,\,(\mathbf{b}_{pol}=b_r\hat{r}+b_{\theta}\hat{\theta},\,\mathbf{b}_{tor}=b_{\phi}\hat{\phi},\,\mathbf{u}_{tor}=u_{\phi}\hat{\phi})\nonumber
\end{eqnarray}
\vspace*{2pt}
\end{figure*}

\subsection{Comparison of Field Structure model and $\alpha\Omega$ dynamo}
Now we extended the local 2D cartesian coordinate system to a spherical 3D coordinate system in a limited way. We set $\hat{z}$ to be a toroidal unit vector ($\rightarrow\hat{\mathbf{\phi}}$) and $\hat{y}$ (or $\hat{x}$) to be a poloidal one ($\rightarrow\hat{r},\,\hat{\mathbf{\theta}}$). We also assume that a differential rotation effect generates a toroidal velocity field: $\mathbf{u}_{tor}=\mathbf{\Omega} \times \mathbf{r}=r\Omega(r,\,\theta)\,\mathrm{sin}\,\theta \, \hat{\phi}$ where `$\mathbf{\Omega}$' is the angular velocity.\\

\noindent In the field structure model if $\mathbf{u}_z\cdot\nabla \mathbf{b}_y$ is $zero$ ($\sim$ axisymmetric) or $positive$, $\mathbf{b}_y$ cannot be amplified without an additional forcing source. But if $\mathbf{u}_z\cdot\nabla \mathbf{b}_y$ and $\mathbf{b}_y\cdot\nabla \mathbf{u}_z$ have a nontrivial negative and positive value respectively, $\mathbf{b}_y$ and $\mathbf{b}_z$ can induce each other self consistently. The converted Eq.~(\ref{poloidal field alpha-Omega dynamo}), (\ref{toroidal field alpha-Omega dynamo}) in a spherical coordinate system show the same feature in a more practical way. If the spherical system is not axisymmetric, i.e., weakly axisymmetric, $\mathbf{b}_{pol}$ can be driven by `$\mathbf{1a}$' in Eq.~(\ref{poloidal field alpha-Omega dynamo}) ($-\Omega(\frac{\partial b_r}{\partial \phi}\hat{r}+\frac{\partial b_{\theta}}{\partial \phi}\hat{\theta})$). But if the system is axisymmetric, `$\mathbf{1a}$' disappears so that $\mathbf{b}_{pol}$ decays without a forcing source. On the contrary, $\mathbf{b}_{tor}$ is driven by the term `$\mathbf{b}_{pol}\cdot \nabla \mathbf{u}_{tor}$' which is related to the differential rotation and by the term $-\Omega(b_r\frac{\partial \hat{r}}{\partial \phi}+b_{\theta}\frac{\partial \hat{\theta}}{\partial \phi})$ ($\mathbf{1b}$, Eq.~(\ref{toroidal field alpha-Omega dynamo})) whether the system is axisymmetric or not. They are merged into a well known forcing source in the $\alpha\,\Omega$ dynamo \citep{Charbonneau 2013}:
\begin{eqnarray}
-\Omega(b_r \,\mathrm{sin}\,\theta+b_\theta \, \mathrm{cos}\,\theta)+b_r \,\frac{\partial }{\partial r}(\Omega\,r\,\mathrm{sin}\,\theta) + \frac{b_\theta}{r}\frac{\partial }{\partial \theta}\big(\Omega\,r\,\mathrm{sin}\theta\big)\nonumber\\
\rightarrow r\,\mathrm{sin}\theta\,(\mathbf{b}_{pol}\cdot\nabla\Omega).
\label{toroidal and poloidal field alpha-Omega dynamo}
\end{eqnarray}
This simple comparison clearly shows that an exact axisymmetric system cannot sustain $\mathbf{b}_{pol}$ and $\mathbf{b}_{tor}$ self-consistently \citep{Cowling 1934}. It also implies one of the conditions to break the axisymmetry. The initially given energy ($u_{z}\partial b_y/\partial z<0$) due to some reason can break the symmetry so that $\mathbf{b}_{pol}$ and $\mathbf{b}_{tor}$ drive each other leading to the self consistent dynamo process.


\section{Conclusion}
We have discussed the inverse transfer of helical and nonhelical magnetic energy $E_M$ in a free decaying MHD system. In case of a helical MHD system, $E_M$ is inversely cascaded due to the $\alpha$ effect whether the system is forced or not. The simulation and analytic Eqs.~(\ref{HmSolution1}), (\ref{EmSolution2}) confirm this and explain how the large scale magnetic field grows before decay in the non forcing (decaying) MHD system. For a nonhelical decaying MHD system, there is no appropriate theory yet to explain the evolution of energy across differential spatial scales as shown in the simulation. To approach the problem, we first introduced the conventional methods like the statistical approach and scaling invariant method to discuss about the general principle of magnetic energy transfer. The basic mechanism of energy migration or its distribution is related to the conservation of system variables and the intrinsic properties of scaling invariant MHD equations. However, since these conventional theories are not suitable for a decaying MHD system, we suggest a field structure model based on the magnetic induction equation $\partial \mathbf{b}/\partial t \sim \nabla \times (\mathbf{u}\times \mathbf{b})$. The model shows that the transport of $E_M$ is essentially bi-directional. But the energy transport is constrained by several factors such as local inhomogeneity in the velocity and magnetic field, resulting in nonzero values for $\nabla \mathbf{u}$ and $\nabla \mathbf{b}$, eddy turnover (energy cascade) time $\tau$, helicity, instability, and so on. When a MHD system is mechanically and nonhelically driven, the thermal pressure $-\nabla p$ and the advection term $-\textbf{u}\cdot\nabla \textbf{u}$ transfer the kinetic energy $E_V$ chiefly toward the dissipation scale resulting in the decrease of $\tau$ on small scales. This accelerates the migration of $E_V$ toward the dissipation scale. Then $E_M$ mainly interacts with the forward cascaded $E_V$ to generate new $E_M$ in the small scale regime. However, without the external forcing source the energy in the small scale regime decays fast so that $\tau$ in the small scale elevates more quickly than that of the large scale lowering the efficiency of the forward cascade of $E_V$. In contrast, since $\tau$ in the large scale regime, which has relatively less dissipation effect than small scale regime, does not grow so fast. $E_V$ is more easily led to the large scale. Interacting with the present $E_M$ in the injection or small scale, $E_V$ induces new $E_M$ in large scale regime, i.e., the inverse transfer of magnetic energy in a decaying system. We also have shown the possibility of inverse and forward transfer of $E_M$ using EDQNM approximation in a qualitative way.\\

\noindent Also we compared the simplified magnetic induction equation from the 2D cartesian field structure and from the 3D spherical coordinate system. When the system, forced by the differential rotation effect, is axisymmetric, there is no internal term that drives the poloidal magnetic field $\mathbf{b}_{pol}$. This makes the self-consistent dynamo process impossible, which means the decay of $\mathbf{b}_{pol}$. However, if the axisymmetry is broken, the differentiation of the $|{b}_{pol}|$ over the azimuthal angle $\phi$ has a nontrivial value so that $\mathbf{b}_{pol}$ and $\mathbf{b}_{tor}$ can drive each other. A simple analytic equation based on the field structure shows how the initially given energy can make a self-consistent dynamo process possible.\\

\noindent The helical magnetic fields are potentially relevant not only for the energy inverse cascade, but have also been observed in star-forming regions like Orion \citep{Heiles 1997, Stutz and Gould 2016}. The investigation of the implications of such fields thus becomes more related from a theoretical perspective to the actual astrophysical phenomena (see e.g. Fiege and Pudritz 2000a; Fiege and Pudritz 2000b; Gellert et al 2011). Explaining their origin, for instance due to an inverse cascade as examined here, may thus become increasingly relevant in the future.\\

\noindent Finally we can consider the inverse transfer of nonhelical $E_M$ in a kinetically dominated MHD system as a next project. If the system is driven by nonhelical kinetic energy, most of $E_V$ flows toward the small scale regime due to the advection term, pressure, and decreasing eddy turnover time ($\tau\sim 1/ku$). Then $E_M$ is more easily induced in the small scale than in the large scale. Partial $E_M$ can be generated in the larger scale, but will not be so much. The cascaded energy in the small scale regime disappears fast due to the increasing dissipation effect $\sim k^2 E^{1/2}$. So the overall inverse transfer of $E_M$ will decrease compared with magnetically dominated case. However, since it is quite probable that an ephemeral astrophysical event can provide the system with kinetic and electromagnetic energy together, it also deserves to study the migration of $E_M$ in a kinetically dominated MHD system.\\

\section*{Acknowledgements}
KWP appreciates the support from ERC Advanced Grant STARLIGHT: Formation of the First Stars (contract 339177).  KWP also appreciates the discussion with Dr. Ralf Klessen and Dr. Dominik Schleicher.

\label{lastpage}

\begin{thebibliography}{99}
\bibitem[Balbus and Hawley (1991)]{Balbus and Hawley 1991}
Balbus, S. A. and Hawley, J. F., 1991, \apj, 376, 214B

\bibitem[Biermann (1950)]{Biermann 1950}
Biermann, L. 1950, Z. Naturforsch., 5, 65B

\bibitem[Biskamp \& M\"uller (1999)]{Biskamp and Mueller 1999}
Biskamp, D., M\"uller, W., C., 1999, Phys. Rev. Lett., 83, 2195

\bibitem[Biskamp (2008)]{Biskamp 2008}
Biskamp, D., Magnetohydrodynamic Turbulence, 2008, Cambridge University Press, UK

\bibitem[Blackman and Field (2002)]{Blackman and Field 2002}
Blackman E. G., Field G. B., 2002, Phys. Rev. Lett., 89, 265007



\bibitem[Brandenburg (2001)]{Brandenburg 2001}
Brandenburg, A., 2001, ApJ, 550, 824

\bibitem[Brandenburg and Subramanian (2005)]{Brandenburg and Subramanian 2005}
Brandenburg A., Subramanian K., 2005, Phys. Rep., 417, 1

\bibitem[Brandenburg et al (2015)]{Brandenburg et al 2015}
Brandenburg, A., Kahniashvili, T., \& Tevzadze, A., G., 2015, Phys. Rev. L, 114, 075001

\bibitem[Brandenburg \& Kahniashvili (2017)]{Brandenburg and Kahniashvili 2017}
Brandenburg, A., Kahniashvili, T., 2017, Phys. Rev. L, 118, 055102


\bibitem[Charbonneau (2013)]{Charbonneau 2013}
Charbonneau, P., 2013, Solar and Stellar Dynamos: Saas-Fee Advanced Course, 39

\bibitem[Christensson et al. (2001)]{Christensson et al 2001}
Christensson, M., Hindmarsh, M., and Brandenburg, A., 2001, Phys. Rev. E., 64, 056405

\bibitem[Cowling (1934)]{Cowling 1934}
Cowling, T. G., 1934, MNRAS, 94, 768


\bibitem[Davidson (2004)]{Davidson 2004}
Davidson, P., A., An introduction for scientists and engineers, 2004, Oxford University Press

\bibitem[Ditlevsen et al. (2004)]{Ditlevsen et al. 2004}
Ditlevsen, P., D., Jensen, M., H., and Olesen, P., 2004, Phys. A, 342, 471

\bibitem[Fiege and Pudritz (2000a)]{Fiege and Pudritz 2000a}
Fiege, J., D., \& Pudritz, R., E., 2000, MNRAS, 311, 85

\bibitem[Fiege and Pudritz (2000b)]{Fiege and Pudritz 2000b}
Fiege, J., D., \& Pudritz, R., E., 2000, MNRAS, 311, 105

\bibitem[Federrath et al (2011)]{Federrath et al 2011}
Federrath, C., Chabrier, G. , Schober, J.,  Banerjee, R., Klessen, R., S., and Schleicher, D., R., G., 2011,
Phys. Rev. Lett. 107, 114504

\bibitem[Fyfe and Montgomery (1976)]{Fyfe and Montgomery 1976}
Fyfe, D. and Montgomery, D., 1976, J. Plasma Phy., 16, 181

\bibitem[Frisch et al. (1975)]{Frisch et al. 1975}
Frisch, U., Pouquet, A., L$\acute{\mathrm{e}}$orat, 1975, J., Mazure, A. J. Fluid Mech. 68, 769-778

\bibitem[Gellert et al. (2011)]{Gellert et al 2011}
Gellert, M., R{\"u}diger, G., and Hollerbach, R., 2011, MNRAS, 414, 2696G

\bibitem[Harrison (1970)]{Harrison 1970}
Harrison, E., R., 1970, MNRAS, 147, 279.

\bibitem[Haugen et al (2004)]{Haugen et al 2004}
Haugen, N., E., L., Brandenburg, A., and Mee, A., J., 2004, MNRAS, 353, 947

\bibitem[Heiles (1997)]{Heiles 1997}
Heiles, C., 1997, ApJS, 111, 245

\bibitem[Goldstein et al (2013)]{Goldstein et al 2013}
Goldstein, H., Poole, C., P., and Safko, J., L., 2013, Prentice Hall International 3rd ed.

\bibitem[Grasso and Rubinstein (2001)]{Grasso and Rubinstein 2001}
Grasso, D. \& Rubinstein, H. R., 2001, Phys. Rep., 348, 163


\bibitem[Kahniashvili et al. (2013)]{Kahniashvili et al 2013}
Kahniashvili, T., Tevzadze, A. G., Brandenburg, A., \& Neronov, A., 2013, Phys. Rev. D, 87, 083007

\bibitem[Kazantsev (1968)]{Kazantsev 1968}
Kazantsev, A., P., 1968, JETP, 26, 1031

\bibitem[Kraichnan and Nagarajan (1967)]{Kraichnan and Nagarajan 1967}
Kraichnan R. H. \& Nagarajan, S., 1967, Physics of Fluids, 10, 859

\bibitem[Krause and R\"adler (1980)]{Krause and Radler 1980}
Krause, F. \& R\"adler, K. H., 1980, Mean-field magnetohydrodynamics and dynamo theory

\bibitem[Kulsrud and Anderson (1992)]{Kulsrud and Anderson 1992}
Kulsrud, R. M. \& Anderson, 1992, S. W., ApJ, 396, 606

\bibitem[Kulsrud (1999)]{Kulsrud 1999}
Kulsrud, R. M. 1999, ARA \& A, 37, 37


\bibitem[Latif \& Schleicher (2016)]{Latif and Schleicher 2016}
Latif, M., A., and Schleicher, D., R., G., 2016, Astron. Astrophys., 585, 151L

\bibitem[McComb(1990)]{McComb 1990}
McComb, W. D., 1990, The physics of fluid turbulence

\bibitem[Moffatt (1980)]{Moffatt 1980}
Moffatt, H., K., 1980, Magnetic field generation in electrically conducting fluids

\bibitem[Montgomery \& Turner (1981)]{Montgomery and Turner 1981}
Montgomery, D., and Turner, L., 1981, Phys. Fluids, 24, 825-831

\bibitem[Olesen (1997)]{Olesen 1997}
Oleson, P., 1997, PhLB., 398, 321O

\bibitem[Park and Blackman (2012a)]{Park and Blackman 2012a}
Park, K., \& Blackman, E. G., 2012, MNRAS, 419, 913

\bibitem[Park and Blackman (2012b)]{Park and Blackman 2012b}
Park, K., \& Blackman, E. G., 2012, MNRAS, 423, 2120


\bibitem[Park (2014)]{Park 2014}
Park, K., 2014, MNRAS, 444, 3837

\bibitem[Parker (1955)]{Parker 1955}
Parker, E., N., 1955, ApJ, 122, 293

\bibitem[Pouquet et al (1976)]{Pouquet et al 1976}
Pouquet, A., Frisch, U., \& Leorat, J., 1976, J. Fluid Mech., 77, 321

\bibitem[Schekochihin et al. (2002)]{Schekochihin et al 2002}
Schekochihin, A. A., Maron, J. L., Cowley, S. C., \& McWilliams, J. C., 2002, ApJ, 576, 806


\bibitem[Schleicher et al. (2010)]{Schleicher et al 2010}
Schleicher, D., R., G., Banerjee, R., Sur, S., Arshakian, T., G., Klessen, R., S., Beck, R., and Spaans, M., 2010, Astron. Astrophys., 522, 115

\bibitem[Schleicher et al. (2013)]{Schleicher et al 2013}
Schleicher, D., R., G., Schober, J., Federrath, C., and Bovino, S., and Schmidt, W., 2013, New J. Phys., 15, 023017

\bibitem[Schober et al. (2012)]{Schober et al 2012}
Schober, J., Schleicher, D., Bovino, S., and Klessen, R., S., 2012, Phys. Rev. E. 85, 026303

\bibitem[Schmitz (2009)]{Schmitz 2009}
Schmitz, H., A., 2009, ASPC, 413, 98S

\bibitem[Sigl and Olinto (1997)]{Sigl and Olinto 1997}
Sigl, G. \& Olinto, A. V. 1997, Phys. Rev. D., 55, 4582

\bibitem[Stutz and Gould (2016)]{Stutz and Gould 2016}
Stutz, A., M., \& Gould, A., 2016, Astron. Astrophys., 590A, 2S

\bibitem[Subramanian (1998)]{Subramanian 1998}
Subramanian, K., 1998, MNRAS, 294, 718

\bibitem[Subramanian (1999)]{Subramanian 1999}
Subramanian, K., 1999, Phys. Rev. Lett., 83, 2957

\bibitem[Subramanian et al (2014)]{Subramanian and Brandenburg 2014}
Subramanian, K. \& Brandenburg, A., 2014, MNRAS, 445, 2930

\bibitem[Subramanian (2015)]{Subramanian 2015}
Subramanian, K., 2015, arXiv:1504.02311

\bibitem[Turner and Widrow (1988)]{Turner and Widrow 1988}
Turner, M. S. \& Widrow, L. M., 1988, Phys. Rev. D., 37, 2743

\bibitem[Vetterling (2007)]{Vetterling 2007}
Vetterling, W., T., 2007, Numerical Recipes 3rd Edition, Cambridge University Press

\bibitem[Yokoi (2013)]{Yokoi 2013}
Yokoi, N., 2013, Geophys. Astro. Fluid, 107, 114

\bibitem[Yoshizawa(2011)]{Yoshizawa 2011}
Yoshizawa, A., 2011, Hydrodynamic and Magnetohydrodynamic Turbulent Flows: Modelling and Statistical Theory (Fluid Mechanics and Its Applications


\bibitem[Zrake (2014)]{Zrake 2014}
Zrake, J., 2014, ApJL, 794, 26

\end{thebibliography}
\end{document}